\documentstyle[12pt]{report}
\normalbaselines
\newcommand{\cf}[1]{\mbox{\boldmath${#1}$}}
\newcounter{saveeqn}
\newcommand{\alpheqn}{\setcounter{saveeqn}{\value{equation}}%
\setcounter{equation}{0}%
\renewcommand{\theequation}{\mbox{A.\arabic{equation}}}}

\newcommand{\beteqn}{\setcounter{saveeqn}{\value{equation}}%
\setcounter{equation}{0}%
\renewcommand{\theequation}{\mbox{B.\arabic{equation}}}}
\begin{document}
\bibliographystyle{unsrt}
\vbox {\vspace{16mm}} 
\begin{center}
{\Huge \bf  Properties of Squeezed-State\\[4mm] Excitations}\\[7mm]                                         
Vladimir I. Man'ko\\ [3mm]
{\it Lebedev Physical Institute\\
Leninsky Prospekt 53, 117924 Moscow, Russia}\\[3mm]
tel.: (095) 1326219, \quad fax: (095) 9382251, \\[2mm]
e-mail: manko@na.infn.it\\[5mm]
and\\[5mm]
Alfred W\"unsche\\ [5mm]
{\it Arbeitsgruppe "Nichtklassische Strahlung" \\der Max-Planck-Gesellschaft\\
Rudower Chaussee 5, 12489 Berlin, Germany}\\[3mm]
tel.: (4930) 6392 3982, \quad fax: (4930) 6392 3990, \\[2mm]
e-mail: wuensche@photon.fta-berlin.de

\end{center}

\vspace{2mm}

\begin{abstract} 
   The photon distribution function of a discrete series of excitations of
squeezed coherent states is given explicitly in terms of Hermite polynomials
of two variables. The Wigner and the coherent-state quasiprobabilities are
also presented in closed form through the Hermite polynomials and their 
limiting cases. Expectation values of photon numbers and their dispersion are 
calculated. Some three-dimensional plots of photon distributions for 
different squeezing parameters demonstrating oscillatory behaviour are given. 
\\[4mm]
\noindent
PACS number(s): 42.50.Dv, 42.65.Ky

\end{abstract}

\setcounter{chapter}{1}
\setcounter{equation}{0}
\chapter*{1. Introduction}

   The photon and quadrature statistics of nonclassical states of light 
such as squeezed states \cite{wa1,h1}, even and odd coherent states 
\cite{dmm74,y1,buzdj}, displaced Fock ( or number ) states 
\cite{cah,lev,b1,w1} or displaced and squeezed Fock states \cite{kral}, and 
corresponding multimode states as, for example, two-mode squeezed vacuum 
\cite{schrade} differs essentially from the  
statistics of light in coherent states \cite{gla0,gla1,sud1,kla1,buzkn} 
which has the Poissonian photon distribution and Gaussian quadrature 
statistics with equal minimal dispersions of the both noncorrelated 
quadratures. The distributions for nonclassical light have 
frequently oscillatory character \cite{sch1,v1,ag1}. For correlated light 
\cite{dkm1}, this has been found in \cite{kli1} and this phenomenon takes 
place for generalized correlated states \cite{sud2} as well.

In present work, we 
concentrate on statistical properties of states studied in \cite{w2,w3} and 
the aim of the work is to obtain the explicit analytic expressions in terms 
of classical polynomials and in terms of multivariable Hermite polynomials
for the Wigner quasiprobability, the coherent-state quasiprobability and the 
photon distribution function.  The 
discussed states are related to general formulations of different types of 
nonclassical states given in \cite{bial,tit,nie1,spir}. These states form 
a discrete series of excitations of squeezed coherent states in an analogous 
way as the displaced Fock states can be considered as discrete excitations of 
the coherent states. They were introduced for the diagonalization of a class 
of transition operators which make the transition from the density operator
to the quasiprobabilities of the complete Gaussian class by forming the trace. 
For the formulation of orthogonality and completeness relations, it was 
convenient to define these states in a nonnormalized form with a 
parametrization in such a way that in the limit of vanishing squeezing 
parameter they become identical with the displaced Fock states. Other limiting 
cases of these states are the eigenstates of the canonical operators and a 
discrete series of their excitations which has been not considered up to now. 
The squeezed-state excitations are interesting because they combine and
generalize the properties of well-known important states as squeezed states, 
number states, and coherent states.
Such kind of states might appear in the process of parametric excitation of 
the quantum electromagnetic field in resonators with moving walls for which 
nonstationary Casimir effect \cite{l1,m1,d1,j1,jau1} produces squeezing in 
quadrature components both in coherent states and Fock states. 

We will calculate in the present paper the Wigner and the 
coherent-state quasiprobability and furthermore the photon distribution 
function in terms of the multivariable Hermite polynomials \cite{ba1}. 
The distributions and quasidistributions were preliminarily 
given in \cite{w3} in a more conventional form. It was shown that the 
multivariable Hermite polynomials are useful to describe the statistics and 
some other properties of nonclassical states \cite{dmjmp1,djp1,dms,wei1,ma1}. 
One also could realize experimental creation of these states using methods 
of state engineering suggested in \cite{vasch}. 

   It should be remarked that there exists nonclassical light with different
degree of the nonclassicality. For example, the photon statistics of the 
generalized coherent states \cite{bial,tit} and their partial cases like 
Schr\"{o}dinger cats \cite{y1} is the same Poissonian statistics as for usual
coherent states. But the photon quadrature statistics of the generalized 
coherent states is essentially different from the Gaussian statistics of the
uncorrelated quadratures in the coherent states. 
Nonlinear Kerr effect and other nonlinear Hamiltonians depending only on 
photon number operator produce in the process of time evolution from the 
initial coherent states the generalized coherent states and never change
the photon statistics but influence the quadrature statistics of the light,
i.e. the quasiprobabilities. From that point of view, the superposition of
coherent states discussed in
\cite{y1} are closer to classical then Schr\"{o}dinger cats of even and odd 
coherent states \cite{dmm74} in which both photon number and quadrature 
statistics differ from ones in the coherent states. As we will point out the 
quantum statistics of squeezed-states excitations is essentially nonclassical 
in both aspects, i.e. the photon statistics and the quadrature statistics are
essentially different from these statistics in the classical coherent states.
There may be different understandings of notions of classical light. Usually, 
classical light is considered as coherent-state light and the degree of 
deviation from the classical light may be evaluated as the minimal distance
to coherent states \cite{w6} or, in another way, by the minimal parameter $s$
of the $s$-ordered quasiprobabilities for which regions of negativity 
in the quasidistributions begin to appear \cite{lt}.

The paper is organized as follows. In the next section, we consider the 
definitions of the squeezed-state excitations and calculate the normalization
constants of these states in closed form. In section 3, we express the general
products of the squeezed-state excitations in terms of multvariable Hermite 
polynomials. In section 4, we give the explicit expressions of the Wigner
quasiprobability in terms of the multivariable Hermite polynomials and 
discuss some properties of the plots of the quasidistributions. In section 
5, we calculate in closed form the photon  distribution of the squeezed-state
excitations and using plots of the distributions for some sets of parameters
we demonstrate the oscillatory behaviour of the photon distribution. The mean
value of the photon number is found explicitly and discussed in section 6 as
well as dispersions of quadratures and uncertainty product. In two appendices
A and B, the properties of the multivariable Hermite polynomials are reviewed
and some new relations of these polynomials to classical polynomials like 
Legendre, Jacobi, and Gegenbauer polynomials which we used to discuss the 
photon statistics of squeezed-state excitations are derived.

\setcounter{chapter}{2}
\setcounter{equation}{0}
\chapter*{2. Definition of a set of squeezed-state excitations and 
normalization}

A discrete set of squeezed-state excitations $|\beta,n;\zeta\rangle, \;
(\:n=0,1,2,\ldots\:)$ has been introduced in \cite{w2,w3} for the purpose of
a diagonal representation of the complete Gaussian class of quasiprobabilities.
These states possess interesting properties of orthogonality and completeness.
An appropriate way is to introduce these states in three steps. The first step 
is the introduction of squeezed vacuum states with the complex squeezing 
parameter $\zeta$ in the nonunitary approach \cite{w4,w5} and in a 
nonnormalized form according to
\begin{eqnarray}
|0,0;\zeta\rangle &\equiv& (1+|\zeta|^2)^{\frac{1}{4}} \exp\bigg(\!
-\frac{\zeta}{2}\,a^{\dagger\,2}\bigg)|0\rangle \nonumber\\ &=& 
(1+|\zeta|^2)^{\frac{1}{4}}\sum_{m=0}^{\infty}\sqrt{\frac{(2m-1)!!}
{2^m m!}}(-\zeta)^m|2m\rangle,
\end{eqnarray}
with the following scalar product leading to a nonstandard normalization
\begin{equation}
\langle 0,0;\xi|0,0;\zeta \rangle = \Bigg(\frac{(1+|\xi|^2)(1+|\zeta|^2)}
{(1-\zeta \xi^*)^2}\Bigg)^{\!\frac{1}{4}},\quad \langle 0,0;-\zeta| 0,0;
\zeta \rangle = 1.
\end{equation}
The next step is the introduction of a discrete set of excitations of the 
squeezed vacuum states according to
\begin{equation}
|0,n;\zeta\rangle \equiv \frac{1}{\sqrt{n!}}\Bigg(\frac{a^{\dagger}-\zeta^*
a}{\sqrt{1+|\zeta|^2}}\Bigg)^{\! n}|0,0;\zeta\rangle,\quad (\:n=0,1,2,
\ldots\:)\,.
\end{equation}
The last step is the displacement of these states with the unitary 
displacement operator $D(\beta,\beta^*)$ according to
\begin{equation}
|\beta,n;\zeta\rangle \equiv D(\beta,\beta^*)|0,n;\zeta\rangle,\quad
D(\beta,\beta^*) \equiv \exp\big(\beta a^\dagger - \beta^* a\big).
\end{equation}
The states with arbitrary pairs of squeezing parameters $\zeta$ and $-\zeta$ 
and arbitrary displacement parameters $\beta$ are mutually orthonormalized 
and obey a completeness relation as follows 
\begin{equation}
\langle \beta,m;-\zeta|\beta,n;\zeta \rangle = \delta_{m,n},\quad 
\sum_{n=0}^{\infty}|\beta,n;\zeta \rangle \langle \beta,n;-\zeta|=I,
\end{equation}
where $I$ denotes the unity operator of the Fock space.
For vanishing squeezing parameter, one obtains from $|\beta,n;\zeta\rangle$ 
the displaced Fock states $|\beta,n\rangle$
\begin{equation}
|\beta,n;0\rangle \equiv D(\beta,\beta^*)\frac{a^{\dagger\,n}}{\sqrt{n!}}|0\rangle =
D(\beta,\beta^*)|n\rangle \equiv |\beta,n\rangle.
\end{equation}
The relations (2.5) form the background for the introduction of the states 
$|\beta,n;\zeta \rangle$ in the described form with a nonstandard 
normalization.
   
   The coordinate representation of the states $|\beta,n;\zeta\rangle$ has
the form \cite{w3} ( $\psi (q;\beta,n;\zeta)\equiv\langle q|\beta,n;\zeta
\rangle$ )
\begin{eqnarray}
\!& & \!\psi (q;\beta,n;\zeta) \nonumber\\
\!&=& \!\frac{1}{\sqrt{2^n n!}}\Bigg(\sqrt{\frac{1+\zeta^*}{1-\zeta}}\:
\Bigg)^{\!n} H_{n}\Bigg(\sqrt{\frac{1+|\zeta|^2}{(1-\zeta)(1+\zeta^*)\hbar}}
\,\bigg(q-\sqrt{\frac{\hbar}{2}}(\beta+\beta^*)\bigg)\Bigg) \nonumber\\ \!& &
\bigg(\frac{1+|\zeta|^2}{(1-\zeta)^2 \hbar \pi}\bigg)^{\!\frac{1}{4}}
\exp\Bigg\{\!-\frac{1+\zeta}{1-\zeta}\frac{1}{2\hbar}
\bigg(q-\sqrt{\frac{\hbar}{2}}(\beta+\beta^*)\bigg)^{\! 2}+\frac{(\beta-
\beta^*)q}{\sqrt{2\hbar}}-\frac{\beta^2-\beta^{*\,2}}{4} \Bigg\},\nonumber\\
\!& &
\end{eqnarray}
where $H_{n}(z)$ denotes the Hermite polynomials. This formula corresponds
to the following connection between the boson operators $a$ and $a^\dagger$
and the canonical operators $Q$ and $P$
\begin{equation}
a\equiv\frac{Q+iP}{\sqrt{2\hbar}},\quad a^\dagger\equiv\frac{Q-iP}{\sqrt{2\hbar}},
\quad i\hbar[a,a^\dagger]=[Q,P]=i\hbar I,
\end{equation}
and can be calculated using $\langle q|Q=q\langle q|$ and $\langle q|P=
-i\hbar\frac{\partial}{\partial q}\langle q|$ and the normalization 
$\langle q| q^\prime \rangle = \delta(q-q^\prime)$. The momentum 
representation of the states $|\beta,n;\zeta\rangle$ will be given in
section 3.

   In this section, we recalculate the normalization constant $N_{n}(|\zeta|)$ 
of the states $|\beta,n;\zeta\rangle$ given in \cite{w3} in form of the series
\begin{eqnarray}
N_{n}^{-2}(|\zeta|)&\equiv& \langle 0,n;\zeta|0,n;\zeta \rangle \nonumber\\
&=&\bigg(\frac{1+|\zeta|^2}{1-|\zeta|^2}\bigg)^{\!n+
\frac{1}{2}}\sum_{k=0}^{[n/2]}\frac{n!}{k!^2 (n-2k)!}\bigg(\frac{|\zeta|}{1+
|\zeta|^2}\bigg)^{\!2k}.
\end{eqnarray}
This series may be either reorganized or recalculated in the form of the
integral
\begin{equation}
N_{n}^{-2}(|\zeta|)=\int\limits_{-\infty}^{+\infty} dq \:\big|\psi(q;\beta,n;
\zeta)\big|^2,
\end{equation}
which is expressed in terms of Jacobi polynomials $P_{n}^{(j,k)}(z)$ or 
Legendre polynomials $P_{n}(z)$ as follows ( see Appendix A )
\begin{equation}
N_{n}^{-2}(|\zeta|)= \sqrt{\frac{1+|\zeta|^2}{1-|\zeta|^2}} P_{n}^{(0,0)}
\bigg(\frac{1+|\zeta|^2}{1-|\zeta|^2}\bigg) = \sqrt{\frac{1+|\zeta|^2}
{1-|\zeta|^2}} P_{n}\bigg(\frac{1+|\zeta|^2}{1-|\zeta|^2}\bigg).
\end{equation}
A representation by special Gegenbauer polynomials $C_{n}^{\frac{1}{2}}(z)$ 
with upper index $1/2$ is also possible. Thus, for the normalized states  
\begin{equation}
|\beta,n;\zeta\rangle_{norm}\equiv N_{n}(|\zeta|)|\beta,n;\zeta\rangle,
\end{equation}
one obtains in coordinate representation
\begin{eqnarray}
\psi_{norm}(q;\beta,n;\zeta)&\equiv& N_{n}(|\zeta|)\psi(q;\beta,n;\zeta)
\nonumber\\&=& \bigg(\frac{1-|\zeta|^2}{1+|\zeta|^2}\bigg)^{\!\frac{1}{4}}
\bigg\{P_{n}\bigg(\frac{1+|\zeta|^2}{1-|\zeta|^2}\bigg)\bigg\}^{\!-\frac{1}{2}}
\psi(q;\beta,n;\zeta),
\end{eqnarray}
where $\psi(q;\beta,n;\zeta)$ can be taken from Eq.(2.7). The normalization
constants do not depend on the displacement parameter $\beta$ since the
unitary displacement operator is cancelled when forming the scalar 
products as can be seen form Eq.(2.4). The normalization with the 
normalization constant $N_{n}(|\zeta|)$ is only possible for $|\zeta|<1$. 
However, the states $|\beta,n;\zeta \rangle$ are well defined also for 
$|\zeta|\ge 1$ and can be used for auxiliary purposes as, for example, for 
the formulation of completeness relations \cite{w2,w3,w4}.
 
\setcounter{chapter}{3}
\setcounter{equation}{0}
\chapter*{3. Calculation of the general scalar products} 
   In this section, we calculate the general scalar product of two arbitrary 
states $|\alpha,m;\xi\rangle$ and $|\beta,n;\zeta\rangle$. Using Eq.(2.4)
in connection with the product of two arbitrary displacement operators one
can reduce this scalar product to the form
\begin{eqnarray}
\langle \alpha,m;\xi|\beta,n;\zeta\rangle &\equiv& \langle 0,m;\xi|
\big(D(\alpha,\alpha^*)\big)^\dagger D(\beta,\beta^*)|0,n;\zeta\rangle
\nonumber\\ &=& \exp\bigg\{\frac{1}{2}(\alpha^* \beta -\alpha \beta^*)\bigg\}
\langle 0,m;\xi|D(\beta-\alpha,\beta^*-\alpha^*)|0,n;\zeta \rangle\nonumber\\
&\equiv& \exp\bigg\{\frac{1}{2}(\alpha^* \beta -\alpha \beta^*)\bigg\}      
\langle 0,m;\xi|\beta-\alpha,n;\zeta \rangle.
\end{eqnarray}
This formula shows that the general scalar product $\langle \alpha,m;\xi|
\beta,n;\zeta \rangle$ can be obtained from the special scalar product 
$\langle 0,n;\xi|\beta,n;\zeta\rangle$ by the substitution $\beta \rightarrow 
\beta - \alpha$ and multiplication of the result of this substitution by the 
phase factor $\exp\big\{1/2 (\alpha^*\beta-\alpha \beta^*)\big\}$. 

   According to Eq.(2.7) one can calculate the reduced scalar product by 
evaluating the following integral in coordinate representation   
\begin{eqnarray}
& &\langle 0,m;\xi|\beta,n;\zeta\rangle\nonumber\\
&=& \frac{1}{\sqrt{2^{m+n}m!n!}}\bigg(\frac{1+\xi}{1-\xi^*}\bigg)^{\!
\frac{m}{2}}\bigg(\frac{1+\zeta^*}{1-\zeta}\bigg)^{\!\frac{n}{2}}
\frac{\big((1+|\xi|^2)(1+|\zeta|^2)\big)^{\frac{1}{4}}}{\sqrt{(1-\zeta)
(1-\xi^*)\hbar\pi}} \nonumber\\ & & \exp\bigg\{\!-\frac{(\beta+\zeta\beta^*)
(\beta^*+\xi^* \beta)}{2(1-\zeta\xi^*)}\bigg\}A,
\end{eqnarray}
with the following abbreviation for an integral of the Hermite--Gauss type
\begin{eqnarray}
A\!\!&\equiv \!\!&\int\limits_{-\infty}^{+\infty}\! dq\:
H_{m}\Bigg(\sqrt{\frac{1+|\xi|^2}{(1+\xi)(1-\xi^*)\hbar }}\;q\Bigg)
H_{n}\Bigg(\sqrt{\frac{1+|\zeta|^2}{(1-\zeta)(1+\zeta^*)\hbar}}\bigg(q-
\sqrt{\frac{\hbar}{2}}\big(\beta+\beta^*\big)\!\bigg)\!\Bigg)\nonumber\\& &
\!\!\exp\Bigg\{\!-\frac{1-\zeta \xi^*}{(1-\zeta)(1-\xi^*)\hbar}
\bigg(q-\sqrt{\frac{\hbar}{2}}\frac{1-\xi^*}{1-\zeta \xi^*}
(\beta+\zeta \beta^*) \bigg)^{\! 2}\,\Bigg\}.
\end{eqnarray}
With the substitution
\begin{equation}
x= \sqrt{\frac{1+|\xi|^2}{(1+\xi)(1-\xi^*)\hbar }}\;q,
\end{equation}
one obtains an integral of the type of Eq.(B.10) in Appendix B. However,
we wrote the exponential term in the integrand as a Gaussian with displaced 
argument with the consequence that the result of the integration is a pure
polynomial. It can be expressed by the two-variable Hermite polynomials as 
follows 
\begin{eqnarray}
\!\!\!\!&& A = \sqrt{\frac{(1-\zeta)(1-\xi^*)\hbar\pi}{1-\zeta\xi^*}}
H_{mn}^{\{R\}}(y_1,y_2),\nonumber\\ && 
R = \frac{2}{1-\zeta \xi^*}\nonumber\\&& \left( \begin{array}{ccc} 
\displaystyle{\!\!\frac{(1-\xi^*)(\xi+\zeta)}{1+\xi}}\!\!&,&\!\! 
\displaystyle{\!\!-\sqrt{\frac{(1-\zeta)(1-\xi^*)(1+|\zeta|^2)(1+|\xi|^2)}
{(1+\zeta^*)(1+\xi)}}}\\ 
\displaystyle{\!\!-\sqrt{\frac{(1-\zeta)(1-\xi^*)(1+|\zeta|^2)(1+|\xi|^2)}
{(1+\zeta^*)(1+\xi)}}}\!\!&,&\!\!
\displaystyle{\!\!\frac{(1-\zeta)(\xi^*+\zeta^*)}{1+\zeta^*}}\! \end{array} 
\right)\!,
\nonumber\\[3mm] &&
y_1=\sqrt{\frac{(1+\xi)(1+|\xi|^2)}{2(1-\xi^*)}}\frac{\beta^*-\zeta^* \beta}
{1-\xi\zeta^*},\quad y_2=-\sqrt{\frac{(1+\zeta^*)(1+|\zeta|^2)}{2(1-\zeta)}}
\frac{\beta-\xi \beta^*}{1-\xi\zeta^*}.\quad \nonumber\\
\end{eqnarray}
This can be written in terms of the usual Hermite polynomials with the
following final result for the special scalar product in Eq.(3.2)
\begin{eqnarray}
& & \langle 0,m;\xi|\beta,n;\zeta \rangle \nonumber\\   
&=&\Bigg(\frac{(1+|\xi|^2)(1+|\zeta|^2)}{(1-\zeta \xi^*)^2}\Bigg)^{\!
\frac{1}{4}} \exp\bigg\{\!-\frac{(\beta+\zeta \beta^*)(\beta^*
+\xi^* \beta)}{2(1-\zeta \xi^*)}\bigg\}\nonumber\\
& &\frac{(-1)^n}{\sqrt{2^{m+n}m!n!}}
\Bigg(\sqrt{\frac{\xi+\zeta}
{1-\zeta \xi^*}}\;\Bigg)^{\!m} \Bigg(\sqrt{\frac{\xi^*+\zeta^*}{1-\zeta \xi^*}}
\;\Bigg)^{\!n} \:\nonumber\\
& &\sum_{j=0}^{\{m,n \}}\frac{(-1)^j m! n!}{j!(m-j)!(n-j)!}
\Bigg(\frac{2\sqrt{(1+|\xi|^2)(1+|\zeta|^2)}}{|\xi+\zeta|}\Bigg)^j \nonumber\\
& & H_{m-j}\Bigg(\sqrt{\frac{1+|\xi|^2}{2(1-\zeta \xi^*)(\xi+\zeta)}}
(\beta+\zeta \beta^*)\Bigg) H_{n-j}\Bigg(\sqrt{\frac{1+|\zeta|^2}{2(1-\zeta
\xi^*)(\xi^*+\zeta^*)}}(\beta^*+\xi^* \beta)\Bigg).\nonumber\\
\end{eqnarray}
Recall that the general scalar product $\langle \alpha,m;\xi|\beta,n;\zeta 
\rangle$ can be obtained from this special scalar product $\langle 0,m;\xi|
\beta, n;\zeta \rangle$ by the above mentioned simple substitutions ( see
Eq.(3.1) ).

   Let us consider special cases of the general scalar product. The special
case $\xi=\zeta$ is important for the calculation of expectation values for
the states $|\beta,n;\zeta\rangle$ (section 6). In this case, one finds from 
Eq.(3.6) 
\begin{eqnarray}
&& \langle 0,m;\zeta|\beta,n;\zeta \rangle \nonumber\\
&=& \sqrt{\frac{1+|\zeta|^2}{1-|\zeta|^2}}\exp\bigg\{\!-\frac{|\beta + \zeta
\beta^*|^2}{2(1-|\zeta|^2)}\bigg\}\nonumber\\&& \frac{(-1)^n}{\sqrt{m!n!}}
\Bigg(\sqrt{ \frac{\zeta}{1-|\zeta|^2}}\:\Bigg)^{\!m}\Bigg(\sqrt{
\frac{\zeta^*}{1-|\zeta|^2}}\:\Bigg)^{\!n}\sum_{j=0}^{\{m,n\}}
\frac{(-1)^j m!n!}{j!(m-j)!(n-j)!}\bigg(\frac{1+|\zeta|^2}{|\zeta|}
\bigg)^{\!j}\nonumber\\ && H_{m-j}\Bigg(\sqrt{\frac{1+|\zeta|^2}
{(1-|\zeta|^2)\zeta}}\frac{\beta+\zeta \beta^*}{2}\Bigg) H_{n-j}\Bigg( 
\sqrt{\frac{1+|\zeta|^2}{(1-|\zeta|^2)\zeta^*}}\frac{\beta^* + \zeta^* \beta}
{2}\Bigg).
\end{eqnarray}

   In case of equal displacement parameter $\alpha=\beta,$ the scalar product 
$\langle \beta,m;\xi|\beta,n;\zeta \rangle$ becomes independent on $\beta$ 
and can be obtained by setting $\beta=0$ in Eq.(3.6), i.e., by using the 
values of the Hermite polynomials for vanishing argument \cite{w2,w3}
( $H_{n}(0)=\sum_{l=0}^{\infty}(-1)^l (2l)!/l!\delta_{n,2l}$ )
\begin{eqnarray}   
&&\langle \beta,m;\xi|\beta,n;\zeta \rangle \nonumber\\&=& 
\frac{1}{\sqrt{m!n!}}
\Bigg(\sqrt{\frac{1+|\xi|^2}{1-\zeta \xi^*}}\:\Bigg)^{\!m+\frac{1}{2}}
\Bigg(\sqrt{\frac{1+|\zeta|^2}{1-\zeta \xi^*}}\:\Bigg)^{\!n+\frac{1}{2}}
\nonumber\\ && \sum_{k=0}^{[\frac{m}{2}]}\sum_{l=0}^{[\frac{n}{2}]}  
\frac{(-1)^{k+l}m!n}{k!l!\sqrt{(m-2k)!(n-2l)!}}\bigg(\frac{\xi+\zeta}
{2(1+|\xi|^2)}\bigg)^k \bigg(\frac{\xi^*+\zeta^*}{2(1+|\zeta|^2)}\bigg)^l
\delta_{m-2k,n-2l}.\nonumber\\
\end{eqnarray}
This scalar product is nonvanishing only if $|m-n|$ is an even number. By  
setting $m=n+2j$ and $m=n+1+2j,$ one obtains from Eq.(3.7)
\begin{eqnarray}
\!\!\!&&\langle \beta,n+2j;\xi|\beta,n;\zeta \rangle \nonumber\\&=&
\Bigg(\frac{\sqrt{(1+|\xi|^2)(1+|\zeta|^2)}}{1-\zeta\xi^*}\:\Bigg)^{\!n+
\frac{1}{2}}\bigg(\!-\frac{\xi+\zeta}{2(1+|\xi|^2)}\bigg)^j \frac{\sqrt{(n+2j)!
n!}}{(n+j)!}\nonumber\\&& \sum_{l=0}^{[\frac{n}{2}]}\frac{(n+j)!}{l!(l+j)!
(n-2l)!}\bigg(\frac{|\xi+\zeta|^2}{4(1+|\xi|^2)(1+|\zeta|^2)}\bigg)^l
\nonumber\\&=& \bigg(\frac{(1+|\xi|^2)(1+|\zeta|^2)}{(1-\zeta \xi^*)^2}
\bigg)^{\frac{1}{4}}\Bigg(\sqrt{\frac{1-\xi\zeta^*}{1-\zeta\xi^*}}\:
\Bigg)^{\!n} \nonumber\\&& \bigg(\!-\frac{\xi+\zeta}{2(1+|\xi|^2)}\bigg)^j 
\frac{\sqrt{(n+2j)!n!}}{(n+j)!} P_{n}^{(j,j)}
\Bigg(\frac{\sqrt{(1+|\xi|^2)(1+|\zeta|^2)}}{|1-\zeta\xi^*|}\Bigg),
\end{eqnarray}
and
\begin{equation}
\langle \beta,n+1+2j;\xi|\beta,n;\zeta \rangle = 0,
\end{equation}
where we have used the Jacobi polynomials $P_{n}^{(j,k)}(z)$ with $j=k$ for 
the representation ( see Appendices A and B ). In case of $\xi=\zeta,$ one 
finds from Eqs.(3.9) and (3.10) or from (3.7) by setting $\beta=0$
\begin{eqnarray}
&&\langle \beta,n+2j;\zeta|\beta,n;\zeta \rangle \nonumber\\&=& \sqrt{\frac{
1+|\zeta|^2}{1-|\zeta|^2}}\bigg(\!-\frac{\zeta^*}{1-|\zeta|^2}\bigg)^j
\frac{\sqrt{(n+2j)!n!}}{(n+j)!}P_{n}^{(j,j)}\bigg(\frac{1+|\zeta|^2}
{1-|\zeta|^2}\bigg)
\end{eqnarray}
and
\begin{equation}
\langle \beta,n+1+2j;\zeta|\beta,n;\zeta \rangle = 0.  
\end{equation}
The special case $j=0$ of this scalar product gives the inverse squared 
normalization constant of the states $|\beta,n,\zeta \rangle$ which is 
given in Eqs.(2.9) and (2.11). In case of $\xi=-\zeta,$ one finds from 
Eqs.(3.9) and (3.10) the orthogonality relations 
\begin{equation}
\langle \beta,m;-\zeta|\beta,n;\zeta \rangle = \delta_{m,n},
\end{equation}
which are the main reason for the choice of the nonstandard normalization of 
the states $|\beta,n;\zeta\rangle$. 

The states $|\alpha,0;0\rangle$ are 
identical with the coherent states $|\alpha\rangle$. Therefore, one obtains 
the Bargmann representation of the states $|\beta,n;\zeta\rangle$ by an 
analytic function $f(\alpha^*)$ as the following special case of the general
scalar product \cite{w3}
\begin{eqnarray}
f(\alpha^*)&\equiv& \exp\bigg(\frac{|\alpha|^2}{2}\bigg)
\langle \alpha,0;0|\beta,n;\zeta\rangle \nonumber\\&=&
(1+|\zeta|^2)^{\frac{1}{4}}\exp\bigg\{\alpha^* \beta -\frac{\zeta}{2}(\alpha^*
-\beta^*)^2-\frac{|\beta|^2}{2}\bigg\}\nonumber\\& &
\frac{\big(\sqrt{\zeta^*}\,\big)^n}
{\sqrt{2^n n!}} H_{n}\Bigg(\sqrt{\frac{1+|\zeta|^2}{2 \zeta^*}}\:(\alpha^*
-\beta^*\big)\Bigg).
\end{eqnarray}
From this relation, one easily finds the coherent-state quasiprobability
of the states $|\beta,n;\zeta\rangle$. We give it only for $\beta=0$ 
because the transition to arbitrary $\beta$ can be made from $Q(\alpha,
\alpha^*)$ for $\beta=0$ by the simple substitutions $\alpha\: \rightarrow 
\:\alpha-\beta$ and $\alpha^*\:\rightarrow\: \alpha^*-\beta^*$. Thus, one 
obtains for the states $|0,n;\zeta\rangle_{norm}$ taking into account 
$|\alpha,0;0\rangle \equiv |\alpha \rangle $
\begin{eqnarray}
Q(\alpha,\alpha^*)&\equiv& \frac{1}{\pi}\frac{\langle \alpha| 
0,n;\zeta \rangle \langle 0,n;\zeta |\alpha \rangle}{\langle 0,n;\zeta|
0,n;\zeta \rangle }\nonumber\\&=& \frac{\sqrt{1-|\zeta|^2}}{\pi}
\exp \bigg\{\!-\bigg(\alpha \alpha^* + \frac{\zeta^*}{2}\alpha^2 + 
\frac{\zeta}{2}\alpha^{*\,2}\bigg)\bigg\} \nonumber\\ & & \frac{|\zeta|^n}
{2^n n! P_{n} \Big(\frac{1+|\zeta|^2}{1-|\zeta|^2}\Big)} H_{n}\Bigg(\sqrt
{\frac{1+|\zeta|^2}{2\zeta}}\:\alpha\Bigg) H_{n}\Bigg(\sqrt{\frac{1+|\zeta|^2}
{2\zeta^*}}\:\alpha^*\Bigg).
\end{eqnarray}
Figure 1 shows in representation by the canonical variables $(q,p)$ and with
$\hbar=1$ the coherent-state quasiprobability $Q(q,p)$ of the states 
$|0,n;\zeta\rangle_{norm}$ for the first six values $n=0,1,\ldots,5$ 
and for the squeezing parameter $\zeta=(3-\sqrt{5})/2= 0.381966$ from a 
bird's perspective. The plot range is chosen as the maximal range
from $0$ to $1/(2\pi)$ which can be taken on by the normalized coherent-state 
quasiprobability $Q(q,p)$. The maximal height $1/(2\pi)$ is only reached for
coherent states and the square root of the difference of the heigth of
$Q(q,p)$ to $1/(2\pi)$ for the considered state is a measure for its distance 
to a ``classical'' state \cite{w6}. Thus, one can see the nonclassicality of
states in such kind of figures in a very visual way. If one changes the sign 
of the squeezing parameter $\zeta$, in our case by transition to 
$\zeta=-0.381966$, one obtains a quasiprobability $Q(q,p)$ with squeezing 
axes which are rotated about an angle $\pi/2$ relative to the primary 
squeezing axes. The seemingly strange value of the squeezing parameter
$\zeta=(3-\sqrt{5})/2$ was chosen for reason that the asymmetry measure 
$\overline{(\Delta N)^3}$ of the photon number distribution for squeezed 
coherent states takes on the maximal possible negative value that is not
very relevant for the purposes of this paper. The qualitative features of
the pictures for the quasidistributions in the central range of the 
squeezing parameter do not dramatically depend on the value of the 
squeezing parameter.

The position and momentum representation of the states $|\beta,n;\zeta\rangle$
can be obtained from the general scalar product by using that the states 
$|\beta,n;\zeta\rangle$ comprise the states $|q\rangle$ and $|p\rangle$ as   
the following special cases \cite{w2,w3}
\begin{equation}
\bigg|\frac{q+ip}{\sqrt{2\hbar}},0;1\bigg\rangle=(2\hbar \pi)^{\frac{1}{4}}
\exp \bigg(i\frac{pq}{2\hbar}\bigg)|q\rangle,\quad \bigg|\frac{q+ip}{2\hbar},
0;-1\bigg\rangle=(2\hbar \pi)^{\frac{1}{4}}\exp\bigg(\!-i\frac{pq}{2\hbar}
\bigg)|p\rangle.
\end{equation}
In this way, one gets back the position representation of the states 
$|\beta,n;\zeta\rangle$ given in Eq.(2.7) which was the starting point of 
our calculations and in an analogous way we get the following momentum 
representation ( $\psi(p;\beta,n;\zeta)\equiv \langle p|\beta,n;\zeta
\rangle $ )
\begin{eqnarray}
& & \psi(p;\beta,n;\zeta)\nonumber\\ &=& \frac{(-i)^n}{\sqrt{2^n n!}}
\Bigg(\sqrt{\frac{1-\zeta^*}{1+\zeta}}\:\Bigg)^{\!n} H_{n}\Bigg(\sqrt{
\frac{1+|\zeta|^2}{(1+\zeta)(1-\zeta^*)\hbar}}\bigg(p+i\sqrt{\frac{\hbar}{2}}
(\beta-\beta^*)\bigg) \Bigg)\nonumber\\ & & \bigg(\!\frac{1+|\zeta|^2}{(1+ 
\zeta)^2 \hbar\pi}\!\bigg)^{\frac{1}{4}} \exp \bigg\{\!\!-\frac{1-\zeta}
{1+\zeta} \frac{1}{2\hbar}\bigg(\!p+i\sqrt{\frac{\hbar}{2}}(\beta-\beta^*)
\bigg)^2\!-i\frac{(\beta+\beta^*)p}{\sqrt{2\hbar}}
+\frac{\beta^2-\beta^{*\,2}}{4}
\bigg\}.\nonumber\\ 
\end{eqnarray}
The displaced Fock states $|\beta,n\rangle$ are the special cases 
of vanishing squeezing parameter $\zeta$ of the states $|\beta,n;\zeta
\rangle$, i.e. 
\begin{equation}
|\beta,n;0\rangle = D(\beta,\beta^*)\frac{a^{\dagger\,n}}{\sqrt{n!}}|0,0;0
\rangle \equiv |\beta,n\rangle,
\end{equation}
and, therefore, one obtains the scalar products of displaced Fock states 
as special cases of the general scalar products of the states $|\beta,n;\zeta
\rangle$ ( see section 5 ).

\setcounter{chapter}{4}   
\setcounter{equation}{0}
\chapter*{4. Wigner quasiprobability} 
   It is worth to describe the states $|\beta,n;\zeta\rangle$ in terms  
of quasiprobabilities as the Wigner quasiprobability \cite{wig} or the 
coherent-state quasiprobability \cite{hus,kano}. These quasiprobabilities  
have been calculated and given without detailed derivations in \cite{w3}.
In particular, the Wigner quasiprobability was found there in terms of a 
finite series including products of Hermite polynomials. It turns that the
series is summed up, i.e. it is possible to express the Wigner 
quasiprobability using the definition
\begin{equation}
W(q,p)= \frac{1}{2\hbar \pi}\int\limits_{-\infty}^{+\infty}dx\:
\exp\bigg(i\frac{px}
{\hbar}\bigg)\psi\Big(q-\frac{x}{2}\Big)\psi^*\Big(q+\frac{x}{2}\Big),
\end{equation}
by two-variable Hermite polynomials since the integral in Eq.(4.1) is for
the considered states of the same form as calculated in Appendix B. 

The Wigner quasiprobability for the states $|\beta,n;\zeta\rangle$ can be 
obtained from the Wigner quasiprobability for the states $|0,n;\zeta\rangle$
by a displacement $\beta$ of the distribution in the complex plane of the 
complex variable $\alpha$ that means for the real variables $(q,p)$ by the 
substitutions 
\begin{equation}
\alpha\equiv \frac{q+ip}{\sqrt{2\hbar}}\rightarrow\alpha-\beta, \quad
q\rightarrow q-\sqrt{\frac{\hbar}{2}}(\beta+\beta^*),\quad p \rightarrow
p+i\sqrt{\frac{\hbar}{2}}(\beta-\beta^*).
\end{equation}
Therefore, we restrict us to the calculation of the Wigner quasiprobability
for the normalized states $|0,n;\zeta \rangle_{norm}$. From Eq.(4.1) in 
in view of Eqs.(2.7) and (2.12), one finds
\begin{equation}
W(q,p) = \frac{N_{n}^2(|\zeta|)}{2^n n!}\bigg(\frac{|1+\zeta|}{|1-\zeta|}
\bigg)^{\,n} \sqrt{\frac{1+|\zeta|^2}{|1-\zeta|^2\hbar\pi}} 
\exp\bigg\{\!-\frac{|(1+\zeta)q+i(1-\zeta)p|^2}{(1-|\zeta|^2)\hbar}\bigg\}
\frac{1}{2\hbar\pi} \:B,
\end{equation}
with the following abbreviation for an integral of the Hermite--Gauss type
\begin{eqnarray}
B &\equiv& \int\limits_{-\infty}^{+\infty} dx \:
H_{n}\Bigg(\sqrt{\frac{1+|\zeta|^2}{(1-\zeta)(1+\zeta^*)\hbar}}\bigg(
q-\frac{x}{2}\bigg)\Bigg) H_{n}\Bigg(\sqrt{\frac{1+|\zeta|^2}{(1+\zeta)
(1-\zeta^*)\hbar}}\bigg(q+\frac{x}{2}\bigg)\Bigg)\nonumber\\
&&\exp\bigg\{\!-\frac{1-|\zeta|^2}{|1-\zeta|^2\hbar}\bigg(\frac{x}{2}-\frac{
(\zeta-\zeta^*)q+i|1-\zeta|^2\,p}{1-|\zeta|^2}\bigg)^{\!2}\bigg\}.
\end{eqnarray}   
The exponential function in the integrand is here complemented in a way that 
it is a displaced Gaussian function and the result of the integration is a
pure polynomial.
With the substitution
\begin{equation}
z= \sqrt{\frac{1+|\zeta|^2}{(1+\zeta)(1-\zeta^*)\hbar}}
\bigg(q+\frac{x}{2}\bigg),
\end{equation}
one has an integral of the type in Eq.(B.10) in Appendix B. The result in 
terms of two-variable Hermite polynomials is
\begin{eqnarray}
&& B= (-1)^n|1-\zeta|\sqrt{\frac{\hbar\pi}{1-|\zeta|^2}}
H_{nn}^{\{R\}}(y_1,y_2),\nonumber\\&& 
R = \frac{2}{1-|\zeta|^2}
\left(\begin{array}{ccc}
\displaystyle{\frac{2\zeta(1-\zeta^*)}{(1+\zeta)}}&,&
\displaystyle{-\frac{|1-\zeta|(1+|\zeta|^2)}{|1+\zeta|}}\\ 
\displaystyle{-\frac{|1-\zeta|(1+|\zeta|^2)}{|1-\zeta|}}&,& 
\displaystyle{\frac{2\zeta^*(1-\zeta)}{(1+\zeta^*)}} 
\end{array} \right),\nonumber\\&&
y_1=\sqrt{\frac{(1+\zeta)(1+|\zeta|^2)}{(1-\zeta^*)\hbar}}\frac{(1-\zeta^*)q
-i(1+\zeta^*)p}{1-|\zeta|^2},\nonumber\\&&
y_2=-\sqrt{\frac{(1+\zeta^*)(1+|\zeta|^2)}
{(1-\zeta)\hbar}}\frac{(1-\zeta)q+i(1+\zeta)p}{1-|\zeta|^2}.  
\end{eqnarray}
Thus, expression (4.3) contains no series but only well-known special
functions both for the normalization constant and for the 
quadrature-dependent factor.
This can be expressed also in terms of the series for the usual Hermite 
polynomials by 
\begin{eqnarray}
W(q,p)&=&\frac{1}{\hbar\pi}\exp\bigg\{\!-\frac{|(1+\zeta)q+i(1-\zeta)p|^2}
{(1-|\zeta|^2)\hbar}\bigg\}\nonumber\\&&\frac{
\displaystyle{\left(\!-\frac{1+|\zeta|^2}{1-|\zeta|^2}\right)^n}}
{\displaystyle{P_{n}\!\left(\frac{1+|\zeta|^2}{1-|\zeta|^2}\right)}}
\sum_{j=0}^n\frac{(-1)^j n!}{j!^2(n-j)!}\bigg(\frac{
|\zeta|}{1+|\zeta|^2}\bigg)^j \nonumber\\&&
\left|H_{j}\Bigg(\sqrt{\frac{1+|\zeta|^2}{2\zeta(1-|\zeta|^2)\hbar}}
\big((1+\zeta)q+i(1-\zeta)p\big)\Bigg)\right|^{\,2},
\end{eqnarray}
or in complex representation
\begin{eqnarray}
W(\alpha,\alpha^*)&=& \frac{2}{\pi}\exp\bigg\{\!-\frac{(\alpha+\zeta \alpha^*)
(\alpha^*+\zeta^*\alpha)}{1-|\zeta|^2}\bigg\}\nonumber\\&&
\frac{\displaystyle{\left(\!-\frac{1+|\zeta|^2}{1-|\zeta|^2}\right)^n}}
{\displaystyle{P_{n}\!\left(\frac{1+|\zeta|^2}{1-|\zeta|^2}\right)}}
\sum_{j=0}^n \frac{(-1)^j n!}{j!^2(n-j)!}\bigg(\frac{|\zeta|}{1+|\zeta|^2}
\bigg)^j  \nonumber\\&&
H_{j}\Bigg(\sqrt{\frac{1+|\zeta|^2}{\zeta(1-|\zeta|^2)}}(\alpha +\zeta 
\alpha^*)\Bigg) 
H_{j}\Bigg(\sqrt{\frac{1+|\zeta|^2}{\zeta^*(1-|\zeta|^2)}}(\alpha^*+ \zeta^* 
\alpha)\Bigg).
\end{eqnarray}
In representation by the canonical variables $(q,p)$ and with
$\hbar=1,$ Fig.2 shows the Wigner quasiprobability $W(q,p)$ of the states 
$|0,n;\zeta\rangle_{norm}$ for the first six values $n=0,1,\ldots,5$ 
and for the squeezing parameter $\zeta=(3-\sqrt{5})/2=0.381966$ from a frog's 
perspective. The modulus $|\zeta|$ was chosen in such a way that the relation
of the lengths of the major to the minor squeezing axes $x_{max}$ and 
$x_{min}$ which are given by
\begin{equation}
x_{max}=\sqrt{\hbar\,\frac{1+|\zeta|}{1-|\zeta|}},\quad
x_{min}=\sqrt{\hbar\,\frac{1-|\zeta|}{1+|\zeta|}},
\end{equation}
becomes equal to $\sqrt{5}$ ( another reason for this choice was indicated 
in section 3 ). As the plot range is chosen, the maximal possible 
range from $-1/\pi$ to $1/\pi$ which can be taken on by the normalized Wigner 
quasiprobability $W(q,p)$. The maximal values $+1/\pi$ or $-1/\pi$ are
reached at the coordinate origin $q=0,p=0$ for states with even or odd parity, 
respectively, or in more general cases with displacements for states with
definite displaced parity correspondingly defined with respect to this 
displacement. The Wigner quasiprobability $W(q,p)$ of the states 
$|0,n;\zeta\rangle_{norm}$ for the first six values $n=0,1,\ldots,5$ and for 
the squeezing parameter $\zeta=0.5$ from a bird's perspective are given in 
\cite{w2}.

\setcounter{chapter}{5}
\setcounter{equation}{0}
\chapter*{5. Photon statistics of the squeezed-state excitations}
   The photon statistics is described by the photon distribution function
$p_{m}$ which is expressed in terms of the matrix elements $\langle m|\beta,
n;\zeta\rangle$ and taking into account $|m\rangle=|0,m;0\rangle$ by 
\begin{equation}
p_{m}=N_{n}^{2}(|\zeta|)\:|\langle 0,m;0|\beta,n;\zeta \rangle|^2.
\end{equation}
The corresponding matrix elements can be taken either in closed form (3.2), 
(3.5) or as series from Eq.(3.6) as the following special case
\begin{eqnarray}
& & \langle 0,m;0|\beta,n;\zeta\rangle \nonumber\\ &=& (1+|\zeta|^2)^{
\frac{1}{4}}\exp\bigg\{\!-\frac{(\beta+\zeta \beta^*)\beta^*}{2}\bigg\} 
\frac{(-1)^n}{\sqrt{m!n!}}\sum_{j=0}^{\{m,n \}} \frac{(-1)^j m!n!}{j!(m-j)!
(n-j)!} \Big(\sqrt{1+|\zeta|^2}\:\Big)^j \nonumber\\& & \bigg(\frac{\sqrt{
2\zeta}}{2}\:\bigg)^{\! m-j}\bigg(\frac{\sqrt{2\zeta^*}}{2}\:\bigg)^{\! n-j} 
H_{m-j}\bigg(\frac{\beta + \zeta \beta^*}{\sqrt{2\zeta}}
\bigg) H_{n-j}\Bigg(\sqrt{ \frac{1+|\zeta|^2}{2\zeta^*}}\:\beta^*\Bigg),
\end{eqnarray}
whereas the normalization constant $N_{n}(|\zeta|)$ is given by Eqs.(2.9) 
and (2.11).

Let us consider some special cases. For vanishing squeezing parameter
$\zeta,$ one obtains from $|\beta,n;\zeta\rangle$ the displaced Fock states
$|\beta,n;0\rangle \equiv |\beta,n\rangle$ already in the normalized form
( see Eq.(2.6) ) and from Eq.(5.2) it follows with the asymptotic expressions
for the Hermite functions $ H_{n}(z)\rightarrow (2z)^n $
\begin{eqnarray}
\langle 0,m;0|\beta,n;0\rangle &=& \exp\bigg(\!-\frac{|\beta|^2}{2}\bigg)
\frac{(-1)^n}{\sqrt{m!n!}}\sum_{j=0}^{\{m,n\}} \frac{(-1)^j m!n!}{j!(m-j)!
(n-j)!}\beta^{m-j}\beta^{*\,n-j} \nonumber\\&=&\exp\bigg(\!-\frac{|\beta|^2}
{2}\bigg)\sqrt{\frac{n!}{m!}}\beta^{m-n}L_{n}^{m-n}(|\beta|^2)\nonumber \\&=& 
\exp\bigg(\!-\frac{|\beta|^2}{2}\bigg)\sqrt{\frac{m!}{n!}}(-\beta^*)^{n-m}
L_{m}^{n-m}(|\beta|^2), \quad N^{2}_{n}(0)=1,
\end{eqnarray}
where $L_{n}^{\nu}(z)$ denotes the associated Laguerre polynomials ( cf., e.g.,
\cite{w1} ). The photon statistics of displaced Fock states $|\beta;n\rangle$
for $n=0,1,\ldots,5$ and $n=5,6,\ldots,10$ and for the squeezing parameter 
$|\beta|=\sqrt{25/2}=3.53533 $ is shown in Figs.3 and 4. These figures 
demonstrate the influence of the excitation on the photon distribution 
structure. As we see, for larger excitations the structure becomes more  
homogeneous.
  
In the special case of squeezed coherent states $|\beta,0;\zeta\rangle,$ one
obtains from Eq.(5.2)
\begin{eqnarray}
\langle0,m;0|\beta,0;\zeta\rangle \!\!&=&\!\! \big(1
+|\zeta|^2\big)^{\frac{1}{4}}
\exp\bigg\{\!-\frac{(\beta+\zeta \beta^*)\beta^*}{2}\bigg\}
\frac{1}{\sqrt{m!}} \bigg(\frac{\sqrt{2\zeta}}{2}\:\bigg)^{\!m}H_{m}
\bigg(\frac{\beta+\zeta \beta^*}{\sqrt{2\zeta}}\bigg),\nonumber\\
\quad N_{0}^{2}(|\zeta|)\!\! &=&\!\! \bigg(\frac{1-|\zeta|^2}{1+|\zeta|^2}\:
\bigg)^{\frac{1}{2}}.
\end{eqnarray}
In particular, one obtains for squeezed vacuum states $|0,0;\zeta\rangle$ 
\begin{eqnarray}
\langle 0,2m;0|0,0;\zeta\rangle\!\! &=&\!\! \big(1+|\zeta|^2\big)^{\frac{1}{4}}
\frac{\sqrt{2m!}}{2^m m!}(-\zeta)^m,\quad \langle 0,2m+1;0|0,0;\zeta\rangle 
= 0,\nonumber\\  N_{0}^{2}(|\zeta|)\!\! &=&\!\! \bigg(\frac{1-|\zeta|^2}
{1+|\zeta|^2}\:\bigg)^{\frac{1}{2}}.
\end{eqnarray}
The squeezed vacuum states $|0,0;\zeta\rangle$ are superpositions of even
Fock states $|2m\rangle$. The states $|0,1;\zeta\rangle$ are squeezed Fock
states $|1\rangle$ for which one obtains
\begin{eqnarray}
\langle 0,2m+1;0|0,1;\zeta\rangle\!\! &=&\!\! \big(1+|\zeta|^2\big)^{\frac{3}{4}}
\frac{\sqrt{(2m+1)!}}{2^m m!}(-\zeta)^m,\quad \langle 0,2m;0|0,0;\zeta\rangle 
= 0,\nonumber\\  N_{1}^{2}(|\zeta|)\!\! &=&\!\! \bigg(\frac{1-|\zeta|^2}
{1+|\zeta|^2}\:\bigg)^{\frac{3}{2}}. 
\end{eqnarray}
These are superpositions of odd Fock states $|2m+1 \rangle$. More generally,
one obtains for the states $|0,n;\zeta \rangle$
\begin{eqnarray}
&&\langle 0,n+2m;0|0,n;\zeta \rangle \nonumber\\&=&\!\!\Big(\sqrt{1+|\zeta|^2}
\Big)^{n+\frac{1}{2}} \frac{\sqrt{(n+2m)!n!}}{2^m(n+m)!}(-\zeta)^m 
\sum_{l=0}^{[n/2]}\frac{(n+m)!}{l!(l+m)!(n-2l)!}\bigg(\frac{\zeta+\zeta^*}
{4(1+|\zeta|^2)}\bigg)^l \nonumber\\&=&(1+|\zeta|^2)^{\frac{1}{4}}
\frac{\sqrt{(n+2m)!n!}}{2^m(n+m)!}(-\zeta)^m P_{n}^{(m,m)}
\big(\sqrt{1+|\zeta|^2}\big),\qquad m\ge-\bigg[\frac{n}{2}\bigg],\nonumber
\\[2mm] &&\langle 0,n+1+2m;0|0,n;\zeta \rangle = 0, \qquad m\ge-\bigg[
\frac{n+1}{2}\bigg],\nonumber\\ && N_{n}^2(|\zeta|)= \sqrt{\frac{1-|\zeta|^2}
{1+|\zeta|^2}}\bigg(P_{n}\bigg(\frac{1+|\zeta|^2}{1-|\zeta|^2}\bigg)
\bigg)^{-1},
\end{eqnarray}
where $[\nu]$ denotes the integer part of $\nu$. The states $|0,n;\zeta
\rangle$ are only for $n=0$ and $n=1$ squeezed Fock states which can be 
obtained by applying a unitary squeezing operator onto the Fock states
$|0\rangle$ and $|1 \rangle$, respectively, but they cannot be obtained 
for $n\ge2$ by applying a unitary squeezing operator to the corresponding
Fock states $|n\rangle$. This is due to the property that the unitary 
squeezing operators involves squares of the annihilation operator, 
i.e. $a^2$, in the exponent which applied to the Fock states $|n \rangle$ 
with $n\ge2,$ do not annihilate these Fock states and the unitary and the 
nonunitary approach become inequivalent. It is, however, possible to 
determine initial states from which $|0,n;\zeta\rangle$ can be obtained by 
the same unitary squeezing operators as in the cases $n=0$ and $n=1$ from 
$|0\rangle$ and $|1\rangle$ ( see, \cite{w1} ) or one can first act with 
equivalent unitary or nonunitary squeezing operators onto the vacuum state 
and have then to take the corresponding excitation operators as in the given 
definition of the states.

Figures 5, 6, and 7 show the photon statistics of the normalized states 
$|\beta,n;\zeta\rangle$ for $n=0$ and $|\beta|=5$ in dependence on the 
squeezing parameter $\zeta$ in case that the large axis of the squeezing 
ellipse is perpendicular to the displacement parameter $\beta$ 
( $\beta/\sqrt{\zeta}$ real numbers, Figs.5 and 6 ) and parallel to the 
displacement parameter $\beta$ ( $\beta/\sqrt{\zeta}$ imaginary numbers, 
Fig.7 ). The first special case will be called ``perpendicular squeezing''
( $\beta/\sqrt{\zeta}$ real numbers ) and the second special case ``parallel 
squeezing'' ( $\beta/\sqrt{\zeta}$ imaginary numbers ) or we will speak about 
``perpendicular geometry'' and ``parallel geometry,'' respectively. 
Conditionally, one can also speak about ``amplitude squeezing'' and 
``phase squeezing,'' respectively, but this is only correct in a certain 
approximation if the 
quasidistributions are essentially restricted to small sectors of the
phase plane because amplitude squeezing is squeezing in radial direction and
phase squeezing is squeezing in angular direction. The characterization of
the considered special cases by the kind of complex numbers 
$\beta/\sqrt{\zeta}$ is an invariant characterization meaning that it is 
independent of the chosen position of the coordinate system $(q,p)$ in the 
phase plane \cite{w5}. In case of perpendicular geometry, we see the 
oscillations of the Fock-state occupation after the main occupation with 
pulse-like shape for large squeezing parameters $|\zeta|\approx 0.75\ldots 1$. 
This range of squeezing parameters is amplified in Fig.6. The oscillations 
can be explained by the neighbourhood to zeroes of the Hermite polynomials 
for real arguments. Figures 5 and 7 show for vanishing squeezing parameter 
$\zeta$ the photon statistics of coherent states ( Poissonian statistics ). 
Figures 8 and 9 are the analogous pictures for $n=10$ and displacement 
parameter 
$|\beta|=\sqrt{15}=3.87298 $. These photon statistics also show oscillatory 
behaviour where in the second case for an intermediate range of values of 
the squeezing parameter $|\zeta|\approx 0.25\ldots 0.5$ some Fock states 
$|n\rangle$ with low numbers $n$ are occupied and after a gap in the 
distribution appears again a region of occupation with an pulse-like shape.
The cooperation of displacement, squeezing, and excitation leads to a 
diversity of possible features in the photon statistics that must be studied
furthermore.

\setcounter{chapter}{6}
\setcounter{equation}{0}
\chapter*{6. Expection values for the squeezed-state excitations}
   The expection value of an arbitrary operator $A$ for the squeezed-state
excitations $|\beta,n;\zeta \rangle $ is given by
\begin{equation}
\overline{A}\equiv \frac{\langle \beta,n;\zeta|A|\beta,n;\zeta \rangle}
{\langle \beta,n;\zeta|\beta,n;\zeta \rangle} = \frac{\langle 0,n;\zeta|
\big(D(\beta,\beta^*)\big)^\dagger A D(\beta,\beta^*) |0,n;\zeta \rangle}
{\langle 0,n;\zeta|0,n;\zeta \rangle}.
\end{equation}
The scalar product in the denominator determines the normalization factor 
of the states $|\beta,n;\zeta\rangle $ and is expressed by special Jacobi  
or Legendre polynomials in Eq.(2.11). A possible approach for the calculation 
of expectation values of ordered powers of the annihilation and creation
operator, i.e. ordered moments, is the following. First we calculate the
expectation value of the displacement operator $D(\alpha,\alpha^*)$. Due to 
\begin{equation}
\big(D(\beta,\beta^*)\big)^\dagger D(\alpha,\alpha^*) D(\beta,\beta^*)
=\exp(\alpha \beta^*-\alpha^* \beta) D(\alpha,\alpha^*),
\end{equation}
one finds
\begin{equation}
\overline{D(\alpha,\alpha^*)}=\exp(\alpha \beta^* -\alpha^* \beta) \frac{ 
\langle 0,n;\zeta|\alpha,n;\zeta \rangle }{\langle 0,n;\zeta|0,n;\zeta 
\rangle},
\end{equation}
where the scalar product $\langle 0,n;\zeta|\alpha,n;\zeta \rangle $ can be
taken from Eq.(3.7) by setting $m=n$ and by substituting $\beta \rightarrow 
\alpha$. For the well-known identities ( e.g.,\cite{gla1} )
\begin{eqnarray}
&&\exp(\alpha a^\dagger)\exp(-\alpha^* a)=\exp\left(\frac{\alpha \alpha^*}{2}
\right)D(\alpha,\alpha^*), \nonumber\\ &&
\exp(-\alpha^* a)\exp(\alpha a^\dagger)=\exp\left(\!-\frac{
\alpha \alpha^*}{2}\right)D(\alpha,\alpha^*),
\end{eqnarray}
one has
\begin{eqnarray}
&& a^{\dagger\,l}a^k = (-1)^k\left\{ \frac{\partial^{k+l}}{\partial 
\alpha^{*\,k} \partial \alpha^l} \exp\left(\frac{\alpha \alpha^*}{2}\right) 
D(\alpha,\alpha^*)\right\}_{\alpha=\alpha^*=0},\nonumber\\
&& a^k a^{\dagger\,l} = (-1)^k\left\{ \frac{\partial^{k+l}}{\partial 
\alpha^{*\,k} \partial \alpha^l} \exp\left(\!-\frac{\alpha \alpha^*}{2}\right) 
D(\alpha,\alpha^*)\right\}_{\alpha=\alpha^*=0}.
\end{eqnarray}
Hence we find the normally and antinormally ordered moments for the normalized
states $|\beta,n;\zeta \rangle_{norm}$ by
\begin{eqnarray}
\overline{ a^{\dagger\,l} a^k}\!\!&=&\!\!\frac{1}{\langle 0,n;\zeta|0,n;
\zeta \rangle}\nonumber\\ \!\!&& \left\{
\bigg(\beta-\frac{\partial}{\partial \alpha^*} \bigg)^k  
\bigg(\beta^* +\frac{\partial}{\partial \alpha} \bigg)^l 
\exp\left(\frac{\alpha \alpha^*}{2} \right) \langle 0,n;\zeta|\alpha,n;\zeta 
\rangle \right\}_{\alpha=\alpha^*=0},
\nonumber\\
\overline{a^k a^{\dagger\,l}}\!\!&=&\!\!\frac{1}{\langle 0,n;\zeta|0,n;
\zeta \rangle}\nonumber\\ \!\!&& \left\{
\bigg(\beta -\frac{\partial}{\partial \alpha^*} \bigg)^k
\bigg(\beta^* +\frac{\partial}{\partial \alpha} \bigg)^l 
\exp\left(\!-\frac{\alpha \alpha^*}{2} \right) \langle 0,n;\zeta|\alpha,n;
\zeta \rangle \right\}_{\alpha=\alpha^*=0},
\end{eqnarray}
where we have seperated the displacement part $\exp(\alpha \beta^* -\alpha^*
\beta)$ from the differentiations and obtain in such a way by applying the
binomial formula the moments as sums over powers of the displacement 
parameters $\beta$ and $\beta^*$. However, it is not easy to calculate the 
necessary derivatives in these expressions for arbitrary numbers $k$ and $l$. 
They can be expressed by multivariable Hermite polynomials but the transition 
to usual polynomials is then very complicated. Therefore, we represent here a 
second approach which gives the possibility to calculate these ordered moments 
for low order in a more direct way.

   In \cite{w2}, was derived that the operators $a(\zeta)$ and $\big(a(-\zeta)
\big)^\dagger$ defined by
\begin{equation}
a(\zeta)\equiv \frac{a+\zeta a^\dagger}{\sqrt{1+\zeta \zeta^*}},\quad
\big(a(-\zeta)\big)^\dagger \equiv \frac{a^\dagger -\zeta^* a}{\sqrt{1+\zeta
\zeta^*}},\quad \big[a(\zeta),\big(a(-\zeta)\big)^\dagger\big]= I,
\end{equation}
play the role of the annihilation and creation operator of the states $|0,n;
\zeta \rangle$ ( see Eq.(6.17) in \cite{w2} ). With the decomposition
\begin{equation}
a=\frac{a(\zeta)-\zeta \big(a(-\zeta)\big)^\dagger}{\sqrt{1+\zeta \zeta^*}},
\quad a^\dagger=\frac{\big(a(-\zeta)\big)^\dagger+\zeta^*a(\zeta)}
{\sqrt{1+\zeta \zeta^*}},     
\end{equation}
one obtains the action of the operators $a$ and $a^\dagger$ onto the states
$|0,n;\zeta \rangle$ in the form
\begin{eqnarray}
&& a|0,n;\zeta \rangle = \frac{\sqrt{n}|0,n-1;\zeta \rangle -\zeta \sqrt{n+1} 
|0,n+1;\zeta \rangle }{\sqrt{1+\zeta \zeta^*}},\nonumber\\
&& a^\dagger|0,n;\zeta \rangle = \frac{\sqrt{n+1}|0,n+1;\zeta \rangle 
+\zeta^* \sqrt{n} |0,n-1;\zeta \rangle}{\sqrt{1+\zeta \zeta^*}}, 
\end{eqnarray}
and as a consequence, for example, \newpage
\begin{eqnarray}
a^\dagger a|0,n;\zeta \rangle \!\!\!&=&\!\!\! \frac{1}{1+\zeta \zeta^*}
\Big\{(n-(n+1)\zeta\zeta^*)|0,n;\zeta\rangle  \nonumber\\&&\!\!\!
-\zeta \sqrt{(n+2)(n+1)}|0,n+2;\zeta\rangle +\zeta^*\sqrt{n(n-1)}
|0,n-2;\zeta\rangle \Big\},\nonumber\\
a a^\dagger |0,n;\zeta \rangle \!\!\!&=&\!\!\! \frac{1}{1+\zeta \zeta^*}
\Big\{(n+1-n\zeta\zeta^*)|0,n;\zeta \rangle \nonumber\\&&\!\!\!
-\zeta \sqrt{(n+2)(n+1)}|0,n+2;\zeta\rangle +\zeta^*\sqrt{n(n-1)}
|0,n-2;\zeta\rangle \Big\}.
\end{eqnarray}
With
\begin{equation}
(D(\beta,\beta^*))^\dagger a D(\beta,\beta^*) = a+\beta I,\quad
(D(\beta,\beta^*))^\dagger a^\dagger D(\beta,\beta^*) = a^\dagger
+\beta^* I,
\end{equation}
and with the vanishing of the scalar products of the states $|0,n;\zeta 
\rangle$ and $|0,m;\zeta \rangle$ if the difference $|n-m|$ is an odd number,
one easily finds
\begin{equation}
\overline{a}=\beta, \qquad \overline{a^\dagger}=\beta^* .
\end{equation}
In the same way, by using Eq.(6.10) and the explicit expressions for the 
scalar products of states $|0,n;\zeta \rangle$ and $|0,m;\zeta \rangle$ 
( see Eqs.(3.11) and (3.12) ) one finds the expectation value of the number 
operator $N\equiv a^\dagger a$ in the form
\begin{eqnarray}
\overline{N}\!\!&\equiv&\!\!\frac{\langle \beta,n;\zeta|a^\dagger a|\beta,n;
\zeta \rangle}{\langle \beta,n;\zeta|\beta,n;\zeta \rangle} \nonumber\\
\!\!&=&\!\!\frac{\langle 0,n;\zeta|(a^\dagger+\beta^*I)(a+\beta I)
|0,n;\zeta \rangle}{\langle 0,n;\zeta|0,n;\zeta \rangle} \nonumber\\
\!\!&=&\!\!\frac{1}{(1+|\zeta|^2)
\displaystyle{P_{n}^{(0,0)}\left(\frac{1+|\zeta|^2}
{1-|\zeta|^2}\right)}}\Bigg\{\left(n-(n+1)|\zeta|^2\right) P_{n}^{(0,0)}
\left(\frac{1+|\zeta|^2}{1-|\zeta|^2}\right) \nonumber\\ \!\!&&\!\!
+\frac{|\zeta|^2}{1-|\zeta|^2}\Bigg((n+2)
P_{n}^{(1,1)}\left(\frac{1+|\zeta|^2}{1-|\zeta|^2}\right) - n
P_{n-2}^{(1,1)}\left(\frac{1+|\zeta|^2}{1-|\zeta|^2}\right)\Bigg)\Bigg\}
+\beta \beta^*.
\end{eqnarray}
With the following identity for Jacobi polynomials
\begin{equation}
(n+2)P_{n}^{(1,1)}(z)-n P_{n-2}^{(1,1)}(z)=2(2n+1)P_{n}^{(0,0)}(z),
\end{equation}
which can be proved by using the explicit representations of the Jacobi
polynomials with equal upper indices ( e.g., Eq.(A.3) in Appendix A ), 
one obtains from Eq.(6.13) the following simple result of the expectation 
value of the number operator $N\equiv a^\dagger a$ for the normalized states 
$|\beta,n;\zeta\rangle_{norm}$
\begin{equation}
\overline{N}= \frac{n+(n+1)|\zeta|^2}{1-|\zeta|^2}+|\beta|^2.
\end{equation}
In an analogous way, one obtains from
\begin{eqnarray}
a^2|0,n;\zeta \rangle \!\!\!&=&\!\!\! \frac{1}{1+\zeta \zeta^*}
\Big\{\sqrt{n(n-1)}|0,n-2;\zeta \rangle \nonumber\\ &&\!\!\!- \zeta (2n+1)
|0,n;\zeta \rangle + \zeta^2 \sqrt{(n+2)(n+1)}|0,n+2;\zeta \rangle \Big\},
\nonumber\\ 
a^{\dagger\,2}|0,n;\zeta \rangle \!\!\!&=&\!\!\! \frac{1}{1+\zeta \zeta^*}
\Big\{ \sqrt{(n+2)(n+1)}|0,n+2;\zeta \rangle \nonumber\\ &&\!\!\! +\zeta^*        
(2n+1)|0,n;\zeta\rangle +\zeta^{*\,2} \sqrt{n(n-1)}|0,n-2;\zeta \rangle 
\Big\}, 
\end{eqnarray}
by using the representation of the scalar products by means of the Jacobi
polynomials and with the identity in Eq.(6.14)
\begin{eqnarray}
\overline{a^2}\!\!&\equiv&\!\!\frac{\langle \beta,n;\zeta|a^2 
|\beta,n; \zeta \rangle}{\langle \beta,n;\zeta|\beta,n;\zeta \rangle} 
\nonumber\\
&=&\!\! -\frac{\zeta}{1-|\zeta|^2}\left( 2n+1 +n\: 
\frac{\displaystyle{P_{n-2}^{(1,1)}
\left(\frac{1+|\zeta|^2}{1-|\zeta|^2}\right)}}{\displaystyle{
P_{n}^{(0,0)}\left(\frac{1+|\zeta|^2}{1-|\zeta|^2}\right)}}\right)+\beta^2, \qquad
\overline{a^{\dagger\,2}} = {\overline{a^2}}^{\:*}.
\end{eqnarray}
The series of functions $f_{n}(y)$ defined by
\begin{equation}
f_{n}(y)\equiv \frac{\displaystyle{
P_{n-2}^{(1,1)}\left(\frac{1+y}{1-y}\right)}}
{\displaystyle{ P_{n}^{(0,0)}\left(\frac{1+y}{1-y}\right)}}
\end{equation}
has for the first six values of $n$ the following form 
\begin{eqnarray}
&&f_{0}(y)=-(1-y),\quad f_{1}(y)=0, \quad f_{2}(y)=\frac{(1-y)^2}{1+4y+y^2}, 
\quad f_{3}(y)=\frac{2(1-y)^2}{1+8y+y^2},\nonumber\\&& f_{4}(y)=\frac{3(1-y)^2
(1+3y+y^2)}{1+16y+36y^2+16y^3+y^4},\quad f_{5}(y)=\frac{4(1-y)^2(1+5y+y^2)}
{1+24y+76y^2+24y^3+y^4},\nonumber\\
\end{eqnarray}
and gives a contribution in Eq.(6.17) only for $n\ge2$. The uncertainty
of the expectation values of the canonical operators $Q$ and $P$ is given
by
\begin{eqnarray}
&& \overline{(\Delta Q)^2}\equiv \overline{(Q -\overline{Q}I)^2}=
\frac{\hbar}{2}\left\{(2n+1)\frac{|1-\zeta|^2}{1-|\zeta|^2} -\frac{\zeta+
\zeta^*}{1-|\zeta|^2}nf_{n}(|\zeta|^2)\right\}, \nonumber\\
&& \overline{(\Delta P)^2}\equiv \overline{(P -\overline{P}I)^2}=
\frac{\hbar}{2}\left\{(2n+1)\frac{|1+\zeta|^2}{1-|\zeta|^2} +\frac{\zeta+
\zeta^*}{1-|\zeta|^2}nf_{n}(|\zeta|^2)\right\}, 
\end{eqnarray}
and the corresponding symmetrized uncertainty correlation by
\begin{equation}
\overline{\Delta Q \Delta P + \Delta P \Delta Q} = \hbar \frac{i(\zeta -
\zeta^*)}{1-|\zeta|^2}\left(2n+1+nf_{n}\left(|\zeta|^2\right)\right).
\end{equation}
Now it is easy to calculate from (6.20) the uncertainty sum and the 
uncertainty product. The uncertainty sum ( or total uncertainty )
\begin{equation}
\overline{(\Delta Q)^2} + \overline{(\Delta P)^2} =\hbar(2n+1)
\frac{1+|\zeta|^2}{1-|\zeta|^2} \geq 2\, \sqrt{\overline{(\Delta Q)^2}\:
\overline{(\Delta P)^2}}
\end{equation}
is invariant with respect to rotations of the system, i.e. depends not on the
phase of squeezing parameter $\zeta$ but only on its modulus $|\zeta|$ and 
increases linearly with the excitation number $n$ for fixed $|\zeta|$. 
The inequality is simply the relation between arithmetic and geometric means.
The uncertainty product
\begin{eqnarray}
\overline{(\Delta Q)^2}\:\overline{(\Delta P)^2}\!\!&=&\!\! \frac{\hbar^2}{4}
\Bigg\{(2n+1)^2\frac{|1-\zeta^2|^2}{(1-|\zeta|^2)^2} \nonumber\\ &&\!\!
-nf_{n}(|\zeta|^2)\left(2(2n+1)+nf_{n}(|\zeta|^2)\right)\frac{|\zeta+
\zeta^*|^2}{(1-|\zeta|^2)^2} \Bigg\},
\end{eqnarray}
is not rotation-invariant and depends in a complicated manner on the modulus
and on the phase of the squeezing parameter $\zeta$ and increases mainly with 
the excitation number. One can analyse Eqs.(6.20) by a computer for low 
values of the excitation number $n$ and finds that each factor $\overline{
(\Delta Q)^2}$ and $\overline{(\Delta P)^2}$ can be reduced below the value 
$\hbar/2$ for coherent states and reaches for real ( and only real ) positive 
or negative values of $\zeta,$ in the limiting case $\zeta\rightarrow \pm1,$ 
the value zero for $\overline{(\Delta Q)^2}$ or $\overline{(\Delta P)^2}$, 
respectively. Contrary to this, the symmetrized uncertainty correlation 
according to Eq.(6.21) becomes extremal for pure imaginary values of the 
squeezing parameter $\zeta$ under fixed excitation number $n$. 

In an analogous way, one can calculate the expectation values, for example, 
of the operators $a^{\dagger\,2}a^2,\;a^2 a^{\dagger\,2}$ or $N^2$. However,
the expressions obtained are very complicated and one cannot find identities
for the Jacobi polynomials in such a way that the expressions in the 
numerator and in the denominator shortens in an essential way. We do not
write down these complicated expressions but it is possible to find the
effective polynomials in the numerator and denominator for low values of
$n$ in the states $|\beta,n;\zeta \rangle$ by using a computer.

\setcounter{chapter}{7}
\setcounter{equation}{0}
\chapter*{7. Conclusion}

We discussed statistical properties of the squeezed-state excitations both
for the quadrature component of the light and for the photon number. As it
turned out the Wigner quasiprobability and the photon distribution function
for these states are described in terms of multivariable Hermite polynomials. 
The corresponding limiting cases of these multivariable Hermite polynomials 
expressed in terms of Legendre and Laguerre polynomials ( and also 
Gegenbauer and Jacobi polynomials ) gave a possibility both to calculate 
explicitly some series for usual Hermite polynomials and to express the
normalization constant of the squeezed-state excitations and their scalar 
product in terms of the well-studied classical polynomials. Correspondingly,
the coherent-state quasiprobability ( Husimi--Kano function ) was expressed 
in closed form in terms of one-variable Hermite polynomials and Legendre 
polynomials. The influence of displacement parameter and squeezing parameter
onto the photon statistics in the states was calculated. 

It turned out that there exists the range of these parameters for which the
photon distribution function demonstrates highly oscillatory behaviour which
is a characteristics of strongly nonclassical states known for squeezed 
states, even and odd coherent states, and correlated states. The dependence of 
the average photon number on the squeezing and displacement parameter in the
squeezed-state excitations was explicitly given in terms of Jacobi polynomials.
The main results of the work are given in the formulae (2.11), (3.15), (4.3),
(6.15), and (6.20) in which in closed form are described quadrature and photon
statistics of the squeezed-state excitations. Using the method of constructing 
the simple superposition states like even and odd coherent states 
\cite{dmm74} ( or Schr\"odinger cat states \cite{y1} ) it is possible to 
introduce the superposition states in which the partners of the superposition 
are the excitations of the squeezed coherent states introduced in 
\cite{w2,w3} and discussed in the present paper. 

One could conclude that, being dependent on the physical parameters like 
squeezing parameter, displacement parameter, and excitation number parameter,
the squeezed-state excitations demonstrate a reach range of possibilities 
to influence the photon statistics and promise to be realized experimentally 
in future as other members of the nonclassical-state family. \\[2mm]

\newpage

\noindent
{\bf Acknowledgement}\\
One of the authors ( V.I.M. ) thanks the members of the working group 
"Nichtklassische Strahlung" of the Max-Planck Society in Berlin for 
hospitality during the stay.

\alpheqn
\setcounter{chapter}{7}
\setcounter{equation}{0}
\chapter*{Appendix A:\\
New polynomials and their relation to Jacobi, Gegenbauer, and Legendre   
polynomials}

By comparison of two different methods of calculation of scalar products
of the states in Eq.(2.12) for the special case of equal displacement 
parameter we obtained the following relation of new polynomials to the 
Jacobi polynomials $P_{n}^{(i,j)}(z)$ ( see, e.g., \cite{ba1} )
with equal upper indices $i=j$ and with transformed argument $z$ as well as 
multiplied by functions of $z$  
\begin{equation}
\sum_{l=0}^{[n/2]} \frac{(-1)^l (n+j)!}{l!(l+j)!(n-2l)!} \bigg(\frac{x}{4}
\bigg)^{\!l}=\left(\sqrt{1+x}\:\right)^{n}P_{n}^{(j,j)}\bigg(\frac{1}
{\sqrt{1+x}} \bigg), 
\end{equation}
where, according to the parity properties of the Jacobi polynomials, both
signs of the square root $\sqrt{1+x}$ are possible, but one has to take 
the same signs of this square root in the argument of the Jacobi polynomials 
and in the multiplicators. Before proving this relation, we make some
transformations. With the substitution
\begin{equation}
x \equiv -\frac{4y}{(1+y)^2}\,,\quad \leftrightarrow \quad \sqrt{1+x} \equiv 
\frac{1-y}{1+y}\,, \quad \leftrightarrow \quad y \equiv -\frac{\sqrt{1+x}-1}
{\sqrt{1+x}+1}\,,
\end{equation}
one brings Eq.(A.1) into the following form in which we apply it in this 
paper
\begin{equation}
\sum_{l=0}^{[n/2]} \frac{(n+j)!}{l!(l+j)!(n-2l)!} \bigg( \frac{y}{(1+y)^2}
\bigg)^{\!l} = \bigg( \frac{1-y}{1+y} \bigg)^{\!n} P_{n}^{(j,j)} \bigg(
\frac{1+y}{1-y} \bigg).
\end{equation}
Now, by the substitution
\begin{equation}
y \equiv \frac{z-1}{z+1}\,,\quad\leftrightarrow \quad z\equiv\frac{1+y}{1-y}\,,
\end{equation}
one finds from Eq.(A.3)
\begin{equation}
P_{n}^{(j,j)}(z) = z^n \sum_{l=0}^{[n/2]} \frac{(n+j)!}{l!(l+j)!
(n-2l)!} \bigg( \frac{z^2-1}{4z^2} \bigg)^{\!l}. 
\end{equation}

The following relation of the Jacobi polynomials with equal upper indices 
$P_{n}^{(j,j)}(z)$ to the Gegenbauer ( or Ultraspherical ) polynomials 
$C_{n}^{\lambda}(z)$ is known ( \cite{ba1}, chap.10.9, Eq.(4) ) 
\begin{eqnarray}
P_{n}^{(j,j)}(z) \! &=& \! \frac{(n+j)!(2j)!}{(n+2j)!j!} 
C_{n}^{j+\frac{1}{2}}(z) \nonumber\\ \! &=& \! \frac{(n+j)!(2j)!}{(n+2j)!j!
\Gamma\big(j+\frac{1}{2} \big) } \sum_{k=0}^{[n/2]} \frac{ (-1)^k
\Gamma\big(n-k+j+\frac{1}{2}\big)}{k!(n-2k)!}(2z)^{n-2k}, 
\end{eqnarray}
where additionally the explicit representation of the Gegenbauer polynomials
is inserted. There exists another relation between the Jacobi polynomials 
with equal upper indices $P_{l}^{(j,j)}(z)$ and the associated Legendre 
polynomials $P_{n}^{j}(z)$ of the following kind \cite{schrade}
\begin{equation}
P_{n}^{(j,j)}(z) = \frac{(n+j)!}{(n+2j)!}
\bigg(\frac{2}{\sqrt{1-z^2}}\:\bigg)^{j} 
P_{n+j}^{j}(z).
\end{equation}
However, this relation is uniquely determined only in case of even $j=2m$ 
for the two possible signs of the square root $\sqrt{1-z^2}$ and one must be
cautious when applying this relation. The reason is that the associated 
Legendre polynomials $P_{l}^{j}(z)$ with odd upper indices $j$ after the 
substitution $z\equiv \cos\,(\theta)$ do not only depend on $\cos(\theta)$ but 
also depend on $\sin\,(\theta)$. In the particular case $j=0,$ one obtains from 
Eqs.(A.6) and (A.7)
\begin{equation}
P_{n}^{(0,0)}(z) = P_{n}(z) = C_{n}^{\frac{1}{2}}(z),\quad 
\big(\; P_{n}(z) \equiv P_{n}^{0}(z) \;\big).
\end{equation}
This is one case where the relation to the Legendre polynomials is uniquely
defined.

   Let us now prove the given relations. First we prove the transition from
the sum representation of $P_{n}^{(j,j)}(z)$ in Eq.(A.5) to the sum 
representation in Eq.(A.6). By applying the binomial formula to $(1-1/z^2)$ in 
Eq.(A.5), one finds 
\begin{equation}
P_{n}^{(j,j)}(z) = \frac{(n+j)!}{(n+2j)!} \sum_{k=0}^{[n/2]} \frac{(-1)^k}
{k!}z^{n-2k} \sum_{l=k}^{[n/2]} \frac{(n+2j)!}{(l+j)!(l-k)!(n-2l)!2^{2l}}.
\end{equation}
The interior sum over $l$ can be evaluated as follows
\begin{equation}
\sum_{l=k}^{[n/2]}\frac{(n+2j)!}{(l+j)!(l-k)!(n-2l)!2^{2l}} = \frac{
2^{n-2k}(2j)! \Gamma \big(n-k+j+\frac{1}{2}\big)}{j! \Gamma \big(j+\frac{1}{2}
\big)(n-2k)!}.
\end{equation}
This can be proved by complete induction from $n\,\rightarrow\, n+1 $ since 
the result is obviously true for $n=0$ and all integer $j$ and $k \ge 0$.
With the decomposition of the factor $n+1+2j= (n+1-2l)+2(l+j)$ in the 
numerator and the substitution $l \,\rightarrow \, l+1$ in the second 
arising sum term, the proof is easily to perform. Thus, it is proved that the 
expression denoted by $P_{n}^{(j,j)}(z)$ in Eq.(A.5) has the representation 
by the Gegenbauer polynomials given in Eq.(A.6) but it is not yet proved 
by this approach that it is identical with the Jacobi polynomial denoted by 
the same symbol. However, the relations of the Gegenbauer polynomials to 
special Jacobi polynomials are known as already mentioned. Nevertheless, it 
is instructive to follow a direct transition from the explicit general
representation of the Jacobi polynomials specialized for equal upper indices
to the explicit representation by the Gegenbauer polynomials in Eq.(A.6).
We make this next.
   
   Starting from the explicit representation of the Jacobi polynomials with
equal upper indices by applying the binomial formula and the convolution 
formula for the binomial coefficients ( e.g., \cite{w1}, Eqs.(6.6) and (6.7) ), 
one quickly proceeds to the following double sum
\begin{eqnarray}
P_{n}^{(j,j)}(z) \! &=& \! \frac{1}{2^n} \sum_{m=0}^{n} \frac{(n+j)!^2}
{m!(n-m)!(n+j-m)!(j+m)!} \nonumber\\ & & \!
\frac{1}{2} \Big \{(z-1)^{n-m}(z+1)^m+(z+1)^{n-m} (z-1)^m \Big \} 
\nonumber\\ &=& \! \frac{(n+j)!^2}{(n+2j)!2^n} 
\sum_{k=0}^{[n/2]} z^{n-2k}\:\frac{(2n-2k+2j)!}{(n-2k)!} \nonumber\\ & & \!
\sum_{r=0}^{\{2k,n+j\}} \frac{(-1)^r}{r!(2k-r)!(n+j-r)!(n+j-2k+r)!}.
\end{eqnarray}
The interior sum can be evaluated in the following way ( see also \cite{pr}, 
chap.4.2.5, Eq.(29) )
\begin{eqnarray}
\sum_{r=0}^{\{2k,l\}} \frac{(-1)^r}{r!(2k-r)!(l-r)!(l-2k+r)!} \! &=& \! 
\frac{1}{l!(2l-2k)!} \sum_{s=0}^{\{2k,l\}}\frac{(-2)^s(2l-s)!}{s!(2k-s)!
(l-s)!}\nonumber\\ &=& \! \frac{(-1)^k}{k!l!(l-k)!}.
\end{eqnarray}
Using this sum evaluation with the substitution $l = n+j$ in Eq.(A.11) 
together with the ``doubling'' formula for Gamma functions, one obtains
$P_{n}^{(j,j)}(z)$ in the form of Eq.(A.6). The sum evaluation in Eq.(A.12)
can be proved by complete induction from $k \rightarrow k+1$ since it is
true for $k=0$. The sum is transformed in Eq.(A.12) first into another sum. 
This transformation can be made by setting $(-1)^r=(1-2)^r$ and by applying 
the binomial formula to these powers and then after changing the order of 
summations the convolution formula for the binomial coefficients to reduce 
the double sum to a simple sum. It seems that in this transformed
form the proof by complete induction is easier to perform by multiplying
the relation with $(2k+2)(2k+1)$ and by using the decomposition $(2k+2)
(2k+1)=(2k+2-s)(2k+1-s)+2s(2k+2-s)+s(s-1)$. 
In the general case of the Jacobi polynomials $P_{n}^{(i,j)}(z)$ with unequal
upper indices $i\not=j,$ one can make analogous expansions in powers of $z$
( all powers $z^{n-l}$ and not only $z^{n-2k}$ are present in this case ),
where the coefficients are given by finite sums but it was not to
see how these sums can be evaluated by a simple expression as it was possible
and demonstrated for $i=j$. This is an advantage of the direct derivation.

   Last we consider the relation in Eq.(A.7). The Associated Legendre 
polynomials are defined by \cite{dau}
\begin{eqnarray}
P_{l}^{j}(z)\! &\equiv& \! \frac{(-1)^l}{2^l l!} \big(\sqrt{1-z^2}\:\big)^j
\frac{d^{l+j}}{dz^{l+j}}(1-z^2)^{l} \nonumber\\ &=&\! \big(\sqrt{1-z^2}
\:\big)^{j} \frac{1}{2^l } \sum_{k=0}^{[(l-j)/2]}\frac{(-1)^k(2l-2k)!}
{k!(l-j-2k)!(l-k)!}z^{l-j-2k}.
\end{eqnarray}
With the substitution $l=n+j$ and by applying the doubling formula for  
the Gamma function, one finds
\begin{equation}
P_{n+j}^{j}(z) = \big( \sqrt{1-z^2}\:\big)^j \frac{2^j}{\sqrt{\pi}} 
\sum_{k=0}^{[n/2]} \frac{(-1)^k \Gamma \big(n-k+j+\frac{1}{2} \big)} 
{k!(n-2k)!} (2z)^{n-2k}.
\end{equation}
The comparison of this formula with Eq.(A.6) by applying the doubling formula 
of the Gamma function for $(2j)!$ reveals the identity in Eq.(A.7) with the
discussed problems of the sign of the square root.
 
   Despite the close relations of the polynomials in Eq.(A.1) to Jacobi
polynomials with equal upper indices, one must look to them as to new original 
polynomials because these relations include nonlinear argument 
transformations. It seems to be interesting to get more relations for these 
new polynomials as, for example, generating functions, recurrence relations, 
weight factors, and integrals.

\beteqn
\setcounter{chapter}{8}
\chapter*{Appendix B:\\
Multivariable Hermite polynomials}

\noindent

In the Appendix, we review the properties of two-dimensional and
multidimensional Hermite polynomials following \cite{dmjmp1},
\cite{djp1}, \cite{dod89}.

The two-dimensional Hermite polynomials $H_{mn}^{\{{R}\}}
(y_1 ,y_2)$ are defined by means of the generating function \cite{ba1}
\begin{equation}
\exp\Big\{-\frac 12{\cf a}{R}{\cf a}+{\cf a}{R}{\cf y}\Big\}
=\sum_{m,n=0}^{\infty}\frac {a_1^ma_2^n}{m!n!}H_{mn}^{\{{R}\}}({\cf y}).
\end{equation}
Here $a_1$ and $a_2$ are arbitrary complex numbers combined into the
two-dimensional vector ${\cf a}=(a_1,a_2)$ :
\begin{equation}
{\cf a}{R}{\cf a}=\sum_{i,k=0}^2a_iR_{ik}a_k,\qquad {\cf a}
{R}{\cf y}=\sum_{i,k=0}^2a_iR_{ik}y_k,
\end{equation} 
and ${R}$ is the symmetric 2$\times$2--matrix
\begin{equation}
{R}=\left(\begin{array}{clcr}
R_{11}&R_{12}\\
R_{12}&R_{22}\end{array}
\right).
\end{equation}

Introducing the notation
\begin{equation}
r\equiv\frac {R_{12}}{\sqrt {R_{11}R_{22}}},
\end{equation}
one can obtain \cite{dmjmp1} the following formula for the 
two-dimensional Hermite polynomial of zero arguments:  
\begin{equation}
H_{mn}^{\{{R}\}}(0,0)=\mu_{mn}!(-1)^{\frac {m+n}2}
R_{11}^{m/2}R_{22}^{n/2}\left(r^2-1\right)^{\frac {m+n}4}P_{
(m+n)/2}^{|m-n|/2}\left(\frac r{\sqrt {r^2-1}}\right),
\end{equation}
where
\begin{equation}
\mu_{mn}=\mbox{min}\,(m,n),
\end{equation}
and integers $m, n$ must have the same parity ( otherwise the 
right-hand side equals zero ). Here the Hermite polynomial is expressed 
in terms of associated Legendre function.

For coinciding indices, we get 
\begin{equation}
H_{nn}^{\{{R}\}}(0,0)=n!(-\mbox{det}{R})^{\frac 
n2}P_n^{}\left(\frac {-R_{12}}{\sqrt {-\mbox{det}{R}}}\right),
\end{equation}
$P_n(z)$ being the usual Legendre polynomial.  

Equation (B.5) can be rewritten in the following equivalent 
forms, using the Jacobi polynomials 
\begin{equation}
H_{mn}^{\{{R}\}}(0,0)=\frac {m!n!(-1)^{\frac {
m+n}2}}{2^{\frac {|m-n|}2}\left(\frac {m+n}2\right)!}\left(R_{11}^m
R_{22}^n \left(r^2-1\right)^{\mu_{mn}}\right)^{\frac 12}P_{\mu_{m
n}}^{(\frac {|m-n|}2,\frac {|m-n|}2)}\left(\frac r{\sqrt {r^2-1}}\right),
\end{equation}
or using the Gegenbauer polynomials
\begin{equation}
H_{mn}^{\{{R}\}}(0,0)=\frac {\mu_{mn}!|m-n|!(
-1)^{\frac {m+n}2}}{2^{\frac {|m-n|}2}\left(\frac {|m-n|}2\right)
!}\left(R_{11}^m R_{22}^n \left(r^2-1\right)^{\mu_{mn}}\right)^{\frac 
12}C_{\mu_{mn}}^{\frac {|m-n|+1}2}\left(\frac r{\sqrt {r^2-1}}\right).
\end{equation}

For nonzero vector ${\cf y},$ function $H_{mn}^{\{{R}\}}(y_1,y_2)$ 
can be written as a finite sum of products of the usual Hermite 
polynomials, 
\begin{eqnarray}
&&\!\!\left(\frac {R_{11}^mR_{22}^n}{2^{m+n}}\right)^{-\frac 12}H_{mn}^{
\{{R}\}}(y_1,y_2)\nonumber\\
\!\!&=&\!\!\sum_{j=0}^{\mu_{mn}}\left(-\frac {2R_{12}}{\sqrt{
R_{11}R_{22}}}\right)^j \frac {m!n!}{j!(m-j)!(n-j)!}H_{m-j}\left(\frac {
\zeta_1}{\sqrt {2R_{11}}}\right)H_{n-j}\left(\frac {\zeta_2}{\sqrt {
2R_{22}}}\right),\nonumber\\
\end{eqnarray}
where
\begin{equation}
\zeta_1=R_{11}y_1+R_{12}y_2,~~~\zeta_2=R_{12}y_1+R_{22}y_2.
\end{equation}

Using Pauli matrix 
\begin{equation}
\sigma_x=\left(\begin{array}{clcr}
0&1\\ 1&0\end{array} \right),
\end{equation}
one has for nonzero arguments the relation of the Hermite polynomial
to the associated Laguerre polynomial
\begin{equation}
H_{mn}^{\{t\sigma_x\}}(y_1,y_2)=\mu_{mn}!t^{\nu_{
mn}}(-1)^{\mu_{mn}}y_1^{\frac {(n-m+|n-m|)}2}y_2^{\frac {(m-n+|m-n|)}2}
L_{\mu_{mn}}^{|m-n|}(ty_1y_2),
\end{equation}
where
\begin{equation}
\nu_{mn}=\mbox{max}(m,n).
\end{equation}
For zero argument, we have the formula
\begin{equation}
\mbox{$H_{mn}^{\{{R}\}}(0,0)$}=\left(\frac {R_{11}^m R_{22}^n}
{2^{m+n}}\right)^{\frac 12}\sum_{l=0}^{\left[\frac {
\mu_{mn}}2\right]}\frac {(-1)^{\frac {m+n}2}m!n!}
{l!\left(l+\frac{|m-n|}2\right)!\left(\mu_{mn}-2l\right)!}(2r)^{\mu_{mn}-2l},
\end{equation}
where $m$ and $n$ must have the same parity. 

For multivariable Hermite polynomials one can calculate some integrals.
Thus, if we denote
\begin{equation}
{\cf m}=m_{1},m_{2},\ldots ,m_{N},~~~~~{\cf n}=n_{1},n_{2},\ldots
m_{N},~~~~~m_{i},n_{i}=0,1,\ldots\;,
\end{equation}
one has
\begin{equation}
\int d{\cf x}\: 
H_{{\cf m}}^{\{S \}}({\cf x})H_{{\cf n}}^{\{T \}}({\mit\Lambda} {\cf x}
+{\cf d})\exp (-{\cf x}M{\cf x}+{\cf c}{\cf x})
=\frac{\pi^{N/2}}{\sqrt {\det M}}\exp \bigg(\frac {1}{4}{\cf c}
M^{-1}{\cf c}\bigg) H_{{\cf {mn}}}^{\{R \}}({\cf y}),
\end{equation}
where the symmetric 2N$\times $2N--matrix  
\begin{equation}
R=\left( \begin{array}{clcr}R_{11}&R_{12}\\
\widetilde R_{12}&R_{22}\end{array}\right)
\end{equation}
with N$\times $N--blocks $~R_{11},~R_{22},~R_{12}$ is expressed in terms 
of symmetric N$\times $N--matrices $~S,~T,~M$ and N$\times $N--matrix 
$~{\mit\Lambda} $ in the form
\begin{eqnarray}
R_{11}&=&S-\frac {1}{2}SM^{-1}S,\nonumber\\
R_{22}&=&T-\frac {1}{2}T {\mit\Lambda} M^{-1} 
\widetilde {\mit\Lambda} T,\nonumber\\
\widetilde R_{12}&=&-\frac{1}{2}T {\mit\Lambda} M^{-1}S.
\end{eqnarray}
Here the matrix $\widetilde {\mit\Lambda} $ is transposed matrix 
${\mit\Lambda} $, and $\widetilde R_{12}$ is transposed matrix $R_{12}.$ 
The 2N--vector ${\cf y}$ is expressed in terms of N--vectors ${\cf c}$ 
and ${\cf d}$ in the form
\begin{equation}
{\cf y}=R^{-1} \left( \begin{array}{c}{\cf z}_{1}\\
{\cf z}_{2} \end{array} \right ),
\end{equation}
where N--vectors $~{\cf z}_{1}$ and $~{\cf z}_{2}$ are
\begin{eqnarray}
{\cf z}_{1}&=&\frac {1}{4}(SM^{-1}+M^{-1}S){\cf c}\nonumber\\
{\cf z}_{2}&=&\frac {1}{4}(T {\mit\Lambda} M^{-1}
+M^{-1}\widetilde {\mit\Lambda} T){\cf c}+T{\cf d}.
\end{eqnarray}
For matrices $~S=2,~T=2,$ the above formula (B.17) yields
\begin{eqnarray}
&&\!\! \int d{\cf x}\: \bigg\{\prod _{i=1}^{N}H_{m_{i}}(x_{i})H_{n_{i}}
\bigg(\sum _{k=1}^{N} {\mit\Lambda} _{ik}x_{k}+d_{i}\bigg)\bigg\}
\exp (-{\cf x}M{\cf x}+{\cf c}{\cf x})\nonumber\\
&=&\!\! \frac {\pi^{N/2}}{\sqrt {\det M}}
\exp \bigg(\frac {1}{4}{\cf c}M^{-1}{\cf c}\bigg)H_{{\cf {mn}}}^{\{ R \}}
({\cf y}),
\end{eqnarray}
with N$\times $N--blocks $~R_{11},~R_{22},~R_{12}$ expressed in terms of 
N$\times $N--matrices $~M$ and $~{\mit\Lambda} $
in the form
\begin{eqnarray}
R_{11}&=&2(1-M^{-1}),\nonumber\\
R_{22}&=&2(1-{\mit\Lambda} M^{-1}\widetilde {\mit\Lambda} ),\nonumber\\
\widetilde R_{12}&=&-2 {\mit\Lambda} M^{-1}.
\end{eqnarray}
The 2N--vector $~{\cf y}$ is expressed in terms of N--vectors $~{\cf c}$ 
and $~{\cf d}$ in the form of Eq.(B.20) with
\begin{eqnarray}
{\cf z}_{1}&=&M^{-1}{\cf c},\nonumber\\
{\cf z}_{2}&=&\frac {1}{2}( {\mit\Lambda} M^{-1}
+M^{-1}\widetilde {\mit\Lambda} ){\cf c}+2{\cf d},
\end{eqnarray}
according to Eq.(B.21). 

In the special case $N=1,$ the matrices ${\mit\Lambda}$ and $M$ become scalars
and one finds from Eq.(B.22)
\begin{equation}
\int \limits_{-\infty}^{+\infty}dx\:H_{m}(x)H_{n}({\mit\Lambda}x+d)
\exp(-Mx^2+cx) =\sqrt{\frac{\pi}{M}}\exp\bigg(\frac{c^2}{4M}\bigg)
H_{mn}^{\{R\}}(y_1,y_2),
\end{equation}
with 
\begin{equation}
R=2 \left( \begin{array}{ccc}1-\frac{1}{M}&,&-\frac{{\mit\Lambda}}{M}\\[1.2mm]
-\frac{{\mit\Lambda}}{M}&,&1-\frac{{\mit\Lambda^2}}{M} \end{array} \right),
\end{equation}
and
\begin{equation}
\left(\begin{array}{c} z_1 \\[1.2mm] z_2 \end{array} \right) = 
\left(\begin{array}{c} \frac{c}{M} \\[1.2mm] \frac{{\mit\Lambda}c}{M}+2d
\end{array}\right),\qquad
\left(\begin{array}{c} y_1 \\[1.2mm] y_2 \end{array} \right) = 
\frac{1}{2\left(1-\frac{1+{\mit\Lambda^2}}{M}\right)} 
\left( \begin{array}{c}\frac{c}{M}+\frac{2{\mit\Lambda}d}{M}\\[1.2mm]
\frac{{\mit\Lambda}c}{M}-2\left(1-\frac{1}{M}\right)d \end{array} \right). 
\end{equation}
The parameters $\zeta_1$ and $\zeta_2$ in the arguments of the usual Hermite 
polynomials in the representation of the two-variable Hermite polynomials 
according to Eqs.(B.10) and (B.11) are then
\begin{equation}
\zeta_1=z_1,\qquad \zeta_2=z_2.
\end{equation}
Thus, we obtain the following explicit representation of this integral by the 
usual Hermite polynomials   
\begin{eqnarray}
&&\!\!\sqrt{\frac{M}{\pi}}\exp\bigg(-\frac{c^2}{4M}\bigg)
\int \limits_{-\infty}^{+\infty}dx\:H_{m}(x)H_{n}({\mit\Lambda}x+d)
\exp(-Mx^2+cx)\nonumber\\ &=&\!\!
\left(\sqrt{\frac{M-1}{M}}\:\right)^{\!m} \left(\sqrt{\frac{
M-{\mit\Lambda}^2}{M}}\:\right)^{\!n} \sum_{j=0}^{\mu_{mn}}\frac{m!n!}
{j!(m-j)!(n-j)!}\left(\frac{2{\mit\Lambda}}{\sqrt{(M-1)(M-{\mit\Lambda}^2)}}
\right)^j \nonumber\\ &&\!\!H_{m-j}\left(\frac{c}{2\sqrt{M(M-1)}}\right) 
H_{n-j}\left(\frac{{\mit\Lambda}c+2d}{2\sqrt{M(M-{\mit\Lambda}^2)}}\right).
\end{eqnarray}
This means that in the considered special case the calculation of $z_1, z_2,$ 
and $R$ is sufficient for the representation of the result of the 
integral in Eq.(B.25) by sums over the usual Hermite polynomials whereas the 
additional calculation of $y_1$ and $y_2$ is necessary for the representation 
by the two-variable Hermite polynomials.

\newpage

{\LARGE Figure captions}\\[3mm]

\noindent{\large Fig.1}\\[1mm]
Coherent-state quasiprobability $Q(q,p)$ of the squeezed-state excitations 
with the parameters $\beta=0,\,\zeta=0.381966,$ and $n=0,1,2,3,4,5$ 
corresponding to a),b),c),d),e),f) from a bird's perspective ( $\hbar=1$ ). 
\\[1.5mm]
\noindent{\large Fig.2}\\[1mm]
Wigner quasiprobability $W(q,p)$ of the squeezed-state excitations with 
the same parameters $\beta=0,\,\zeta=0.381966,$ and $n=0,1,2,3,4,5$ 
corresponding to a),b),c),d),e),f) as in Fig.1 from a frog's perspective
( $\hbar=1$ ).\\[1.5mm]
\noindent{\large Fig.3}\\[1mm]
Photon distribution function for $\beta=3.53533, \,\zeta=0$ and for 
excitation numbers $ n=0,1,\ldots,5 $, i.e., for displaced Fock states.
\\[1,5mm]
\noindent{\large Fig.4}\\[1mm]
Photon distribution function for $\beta=3.53533, \,\zeta=0$ and for 
excitation numbers $ n=5,6,\ldots,10 $, i.e., for displaced Fock states. 
\\[1.5mm]
\noindent{\large Fig.5}\\[1mm]
Photon distribution function for $\beta=5,\,n=0$ as the function of the  
squeezing parameter $\zeta,\, 0\le \zeta \le 1$, i.e., for squeezed coherent
states and perpendicular geometry ( large squeezing axis perpendicular to
displacement ). \\[1.5mm]
\noindent{\large Fig.6}\\[1mm]
Photon distribution function for $\beta=5,\,n=0$ as the function of the  
squeezing parameter $\zeta,\, 0.75\le \zeta \le 1$, i.e., for squeezed 
coherent states and perpendicular geometry and amplification of the range 
from $0.75$ to $1$. \\[1.5mm]
\noindent{\large Fig.7}\\[1mm]
Photon distribution function for $\beta=5,\,n=0$ as the function of the  
squeezing parameter $\zeta,\, 0\le -\zeta \le 1$, i.e., for squeezed coherent
states and parallel geometry ( large squeezing axis parallel to 
displacement ). \\[1.5mm]
\noindent{\large Fig.8}\\[1mm]
Photon distribution function for $\beta=3.87298,\,n=10$ as the function of 
the squeezing parameter $\zeta,\, 0\le \zeta \le 1$, i.e., for squeezed-state
excitations with perpendicular geometry. \\[1.5mm]
\noindent{\large Fig.9}\\[1mm]
Photon distribution function for $\beta=3.87298,\,n=10$ as the function of 
the squeezing parameter $\zeta,\, 0\le -\zeta \le 1$, i.e., for squeezed-state
excitations with parallel geometry.\\[1.5mm]

\end{document}